\documentclass[12pt]{elsarticle}
\usepackage{graphicx,amsmath,float,tikz, mathtools, bm}
\usepackage[labelformat=simple]{subcaption}

\usepackage[colorlinks=true,linkcolor=blue,citecolor=blue]{hyperref}
\usepackage{setspace}
\usepackage{pgfplots}
\newcommand{\Rey}{Re}
\newcommand{\Pe}{Pe}
\newcommand{\Ca}{Ca}
\newcommand{\muin}{\mu_\text{in}}
\newcommand{\HR}{H_R}
\newcommand{\HC}{H_C}

\newcommand{\nablap}{\nabla_{\mkern-8mu\perp}}

\pgfplotsset{compat=1.18}
\journal{International Journal of Heat and Mass Transfer}

\begin{document}
\begin{frontmatter}
\title{Coupled heat and fluid transport in pulled extrusion of cylinders}

\author[inst1]{Eunice B. Yuwono}

\affiliation[inst1]{organization={School of Computer and Mathematical Sciences and Institute of Photonics and Advanced Sensing},%Department and Organization
            addressline={The University of Adelaide}, 
            %city={City One},
            postcode={5005}, 
            state={SA},
            country={Australia}}

\author[inst1]{Yvonne M. Stokes\texorpdfstring{\corref{cor1}}{}}
\ead{yvonne.stokes@adelaide.edu.au}
\cortext[cor1]{Corresponding author.}
\author[inst2]{Hayden Tronnolone}
\author[inst3,inst4]{Jonathan J. Wylie}

\affiliation[inst2]{organization={College of Science and Engineering},%Department and Organization
            addressline={Flinders University}, 
            %city={Kowloon},
            postcode={5042}, 
            state={SA},
            country={Australia}}
            
\affiliation[inst3]{organization={Department of Mathematics},%Department and Organization
            addressline={City University of Hong Kong}, 
            city={Kowloon},
            % postcode={22222}, 
            % state={State Two},
            country={Hong Kong SAR}}
            
\affiliation[inst4]{organization={Center for Applied Mathematics and Statistics},%Department and Organization
            addressline={New Jersey Institute of Technology}, 
            city={Newark},
            postcode={07102}, 
            state={NJ},
            country={USA}}

\begin{abstract}
 In the fabrication of optical fibres, the viscosity of the glass varies dramatically with temperature so that heat transfer plays an important role in the deformation of the fibre geometry. Surprisingly, for quasi-steady drawing, with measurement of pulling tension, the applied heat can be adjusted to control the tension and temperature modelling is not needed. However, when pulling tension is not measured, a coupled heat and fluid flow model is needed to determine the inputs required for a desired output. In the fast process of drawing a preform to a fibre, heat advection dominates conduction so that heat conduction may be neglected. %Surprisingly, in this case if the output speed is specified, the fibre geometry at the exit of the device is independent of the thermal modelling. 
 By contrast, in the slow process of extruding a preform, heat conduction is important. % and the preform geometry is determined by the viscosity, hence temperature, along its entire length.
 This means that solving the coupled flow and temperature modelling is essential for prediction of preform geometry. %A way of transferring heat, conduction, is not important in fibre drawing but is important in the slower extrusion process. 
 In this paper we derive such a model that incorporates heat conduction for the extensional flow of fibres. The dramatic variations in viscosity with temperature means that this problem is extremely challenging to solve via standard numerical techniques and we therefore develop a novel finite-difference numerical solution method that proves to be highly robust. 
 %Using the assumptions that the length of the thread is much longer than the radius, we show that the asymptotic model is highly nonlinear. The second-order conduction term rises issues in solving the problem. By considering the physics of the problem, this paper utilises an iterative finite difference method to robustly obtain a solution of this boundary-value problem. 
 We use this method to show that conduction significantly affects the size of internal holes at the exit of the device. % in extrusion, and how the second-order extrusion problem connects to the first-order fibre drawing problem. 
\end{abstract}

\begin{keyword}
%% keywords here, in the form: keyword \sep keyword
slender-body theory \sep low-Reynolds-number flows \sep heat transport
% %% PACS codes here, in the form: \PACS code \sep code
% \PACS 0000 \sep 1111
% %% MSC codes here, in the form: \MSC code \sep code
% %% or \MSC[2008] code \sep code (2000 is the default)
% \MSC 0000 \sep 1111
\end{keyword}

\end{frontmatter}

\section{Introduction}
\begin{figure}
    \centering
    \includegraphics[scale = 0.125]{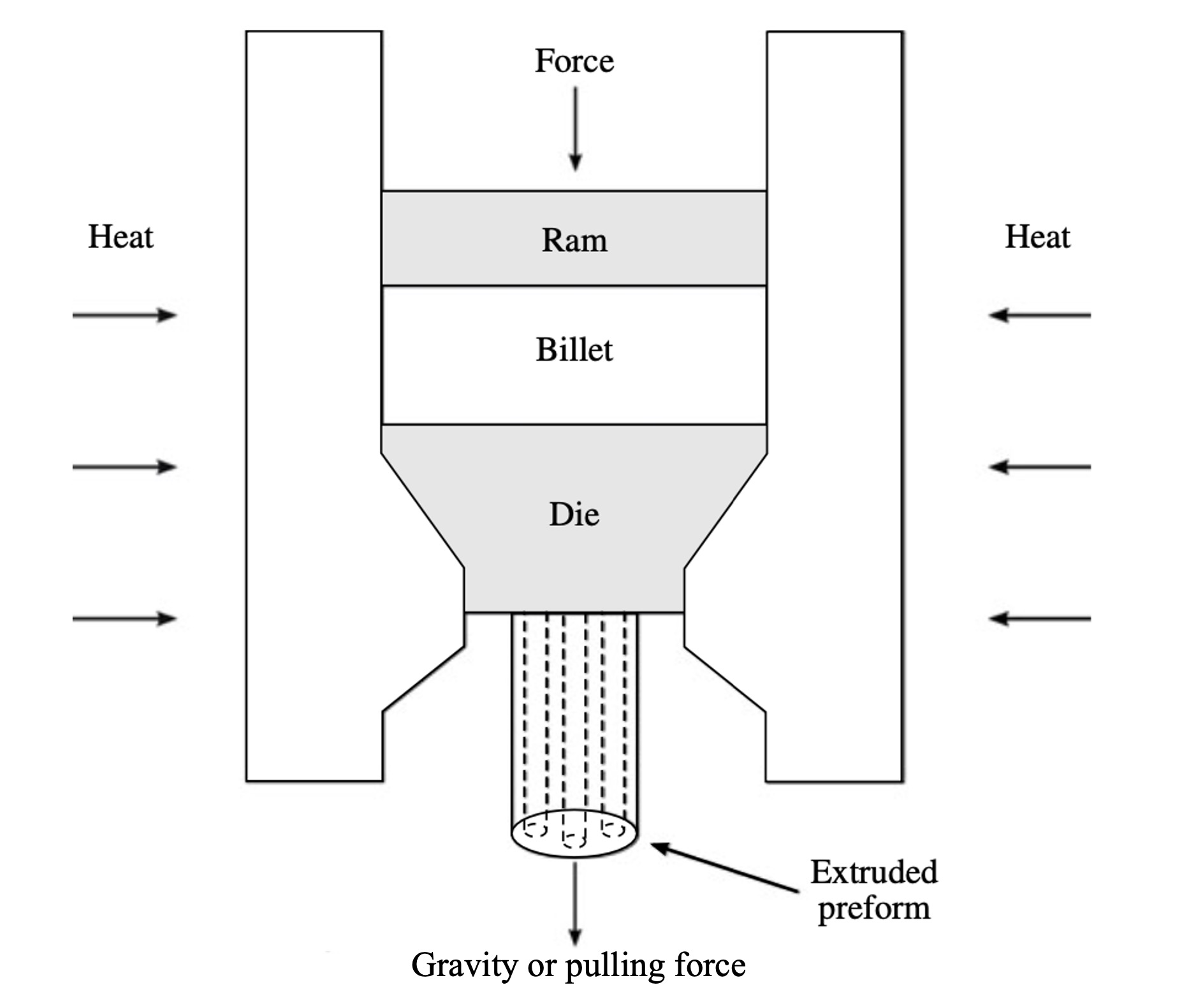}
    \caption{Schematic of extrusion process. The billet is heated and forced through the die by the ram. The blocking elements within the die give rise to the pattern of air channels in the preform. Reproduced from \citet{JFM2016}, [Figure 1].}
    \label{Fig:extrusion_diagram_hayden}
\end{figure}

%\subsection{Microstructured Optical Fibres}
Microstructured optical fibres (MOFs) are optical fibres, typically 100-200 micrometers in diameter, 
%typically made of glass or polymer, 
with one or more internal air channels aligned with the fibre axis. 
%The cross-sections of several different MOFs are shown in Figure \ref{Fig:MOFs}. 
%The diameters of the fibres are typically 100-200 micrometers, and the holes in the cross-section created by the air channels are of the order of micrometres in diameter. 
%Optical fibres guide light, and the internal structure of the air channels affects the optical properties of fibres. By carefully designing the internal structure of MOFs, bespoke optical properties can be obtained that make them extremely valuable in large number of different applications.
%Examples of applications include communication networks \citep{AnnualReview2006}, high-temperature and pressure sensors \citep{HighTempSensing,PressureSensor, TempPressureSensor}, chemical sensing \citep{pH,chemicalmonro, chemical}, and biological sensing such as DNA detection \citep{DNA,DNAsource,  DNA2}.
%The optical properties can depend very sensitively on the geometry of the internal structure and hence it is critical to have very accurate control over the manufacturing process.
% \begin{figure}
%     \centering
%     \includegraphics[scale = 0.7]{the_thesis/chapter1/Figures/1407LFW06f2.jpg}
%     \caption{Cross-sections of some MOFs, as shown in \cite{Laser}. The air channels (dark holes) are aligned with the fibre axis. Photographs are reproduced with the permission of the Institute of Photonics and Advanced Sensing, The University of Adelaide.}
%     \label{Fig:MOFs}
% \end{figure}
%The fabrication of MOFs is a two-step process: preform fabrication and fibre drawing. 
MOF fabrication consists of the fabrication of a macroscopic version of the fibre having a diameter of one or more centimetres, called a fibre preform, followed by the drawing of this preform to a fibre. This paper focuses on fibre-preform fabrication by extrusion, whereby a billet of glass or polymer is heated and pushed through a die by a ram \citep{AnnualReview2006}, as shown in Figure \ref{Fig:extrusion_diagram_hayden}. The die has blockages that result in air channels (or holes) in the extruded preform \citep{Optica}. The extruded preform may simply stretch under its own weight due to gravity or the lower end of the preform may be pulled at a prescribed speed. This paper focuses on pulled extrusion, which is similar to fibre drawing. %The preform, which is a macroscopic version of the fibre with internal structure and diameter of one or more centimetres, is then ``drawn" (i.e. heated and pulled) in the fibre drawing step, reducing the diameter to 100-200 micrometres and producing the final fibre. 
However, in terms of the speed at which the glass moves, preform extrusion is many orders of magnitude slower than fibre drawing. 

One of the main reasons that MOFs have become such an important new technology is that the hole structure can be designed to fabricate fibres that have very special optical properties. This is fundamental in a wide array of applications including communication networks \citep{AnnualReview2006}, high-temperature and pressure sensors \citep{HighTempSensing,PressureSensor, TempPressureSensor}, chemical sensing \citep{pH,chemicalmonro, chemical}, and biological sensing such as DNA detection \citep{DNA,DNAsource,  DNA2}. Nevertheless, the optical properties 
%The optical properties of the fibre, which give a large number of different applications of the MOF, 
can depend very sensitively on the geometry of the internal structure of the fibre and hence achieving very high tolerances in the manufacturing process is paramount. The internal structure undergoes deformation during both the extrusion and drawing stages, and the mechanisms that determine the changes in hole position and shape \citep{Optica} are complicated and hence extremely careful modelling is required. 
%This comprises changes in hole position and changes in shape, and both the holes and the outer boundary can change shape \citep{Optica}.
%In Figure \ref{Fig:extruded_preform}, 
%The holes are created by circular pins in the die, and these holes will deform during the extrusion and drawing process. 
%The changes in the shape and position of the holes affects the structure of the MOF, and consequently affects the optical properties and hence the performance of the fibre. 
%Therefore, we need to choose the initial geometry and conditions of the fibre fabrication process, to adequately control the deformation and give the desired fibre geometry. \\

Temperature is an important factor that affects the deformation of the fibre because of the dramatic changes in viscosity with temperature that occur for the materials that are typically used. %The bulk billet must be heated to enable extrusion, and then the preform is cooled for transfer to the draw tower where it is heated to enable stretching into a fibre. The fibre then must cool before being wound onto the spooling wheel. 
%In both extrusion and drawing, heat is transported by advection and conduction within the glass. In the faster fibre drawing process, advective heat transport is dominant and conduction is assumed to be negligible, but in the case of extrusion conduction is expected to play an important role. %In the case of negligible conduction 
For the quasi-steady processes of fibre drawing and pulled extrusion, where the fibre material enters and exits the deformation region at fixed speeds, mass conservation dictates that the ratio of these speeds determines the change in cross-sectional area. Surprisingly, however, the pulling tension determines % if the speed at the outlet is fixed,
the change in the cross-sectional geometry of the thread \citep{JFM2014}. The pulling tension depends on the temperature but, when this tension is measured, it can be controlled by adjusting the applied heat so that knowledge of the temperature profile in the thread is not needed to achieve a desired geometry change. % are completely independent of the applied heating and cooling, even though other flow properties such as tension in the thread and the velocity profile do depend strongly on the applied temperature. This is an extremely useful result since the paramount property of the fibre from the point of view of applications is precisely the geometry and size of the thread at the exit of the device. 
%Hence, modelling of the applied heating is of limited importance in the case of drawing.
Hence modelling of temperature through the deformation region is of limited importance when pulling tension is measured. However, where pulling tension is not measured (perhaps due to the fragility of the fibre or because the fabrication devices are not suitably equipped), temperature modelling becomes necessary. In the case of fibre drawing, the speed of the process means that advective heat transport is dominant and conduction can be assumed to be negligible \citep{JFM2019}.
%{\color{red} Include here that fibre geometry is determined by harmonic mean of viscosity as given by pulling tension while extrusion requires temperature modelling; see abstract.}
%The fibre drawing process is faster than the preform extrusion process. Therefore, advective heat transport is dominant in drawing and the effect of heat conduction is assumed to be negligible.
By contrast, in the case of pulled preform extrusion (for which tension measurement is is atypical) %temperature modelling is important. Moreover,
%However, in preform extrusion, where 
the process is slower and both advective and conductive heat transport are important. %We will show that the geometry and size of the thread at the exit of the device are significantly affected by the applied temperature and hence that this problem is fundamentally different from the drawing problem.
To understand the effect of conduction in the deformation process, we must carefully model thermal effects to determine the initial geometry and conditions required to obtain a desired output. Developing such a model requires one to overcome a number of challenges including the complicated evolution of free boundaries, possible topological changes if holes close or merge, highly nonlinear dynamics due to dramatic viscosity changes and possible oscillatory instabilities. 

%problem has similar to fibre

%Mathematical modelling is a valuable tool to understand the fibre fabrication process, and the process by which the holes deform. This is because it can reduce costly experimental iterations in order to determine the initial geometry and conditions required. Developing such a model requires one to overcome  a number of challenges: including the complicated evolution of free boundaries, possible topological changes if holes close or merge, highly nonlinear dynamics due to dramatic viscosity changes and possible oscillatory instabilities. \\

The work on preform extrusion is not as extensive as that on fibre drawing. The existing mathematical model of preform extrusion is based on unsteady gravitational extrusion, for which 
%\citet{trabelssi2015} numerically solved a three-dimensional Navier Stokes model of extrusion, neglecting gravity, thermal effects, and surface tension. Incorporating gravity in the arbitrary geometry model, 
\citet{JFM2016, JFM2017} used a similar approach to \citet{JFM2014} to study extrusion of preforms with arbitrary hole geometries. However, their analysis only considers the isothermal case. In contrast to \citet{JFM2016, JFM2017}, this paper explores the quasi-steady pulled extrusion process and couples flow with energy transport. %They employed techniques developed by \cite{wilson1988} in studying dripping from a capillary. In their models the viscosity was assumed to vary with the axial coordinate in a specified way and hence modelling of temperature was not included. \\ 

%\subsection{Literature Review} %remove section %paraphrase section
There has been a significant amount of work done on modelling of the fibre drawing process which is relevant to pulled extrusion. %, but much less on preform extrusion. 
The first model of drawing of axisymmetric solid thread was performed by \citet{Matovich}, assuming isothermal flow. 
%It was extended to an axisymmetric thin walled tube, modelling the film-blowing process as used in manufacture of plastic bags in \cite{pearson1970flow1, pearson1970flow2}. An axisymmetric, isothermal fibre drawing model was developed by \citet{yarin1994}, which included inertial forces, gravity and surface tension. 
\citet{Dew_transport} provided the systematic derivation of an asymptotic model for fibre drawing of non-axisymmetric fibres with isothermal flow. They showed that the cross-sectional shape of fibre changes only in scale in the absence of surface tension. 
%They showed that in the absence of surface tension the cross-sectional shape of the fibre changes only in scale but not in shape. 
This approach was extended to include inertia and gravity by \citet{DHW1994}. Then, \citet{cummingsHowell1999} developed an isothermal, non-axisymmetric model for solid threads with surface tension. This work was extended to model an annular, non-axisymmetric thin walled tube by \citet{griffiths2007, griffiths2008}.

\citet{Fitt2002} modelled the drawing of axisymmetric tubes in a non-isothermal setting. They considered that axial energy transport is dominated by advection, and neglected conduction. Non-isothermal drawing of non-axisymmetric tubes, also neglecting conduction, was considered by \citet{griffiths2008}. %A very good detailed model was explored by 
\citet{Taroni2013} performed a very detailed exploration of non-isothermal drawing %for %applied heating 
of axisymmetric fibres; % drawing. 
they neglected surface tension and also obtained an energy equation that neglects axial conduction. \citet{he2016extension} considered non-isothermal axisymmetric drawing of threads, with temperature-dependent viscosity and surface tension. Again, advection was considered while conduction was neglected. 

The drawing of fibres with arbitrary hole structures was modelled by \citet{JFM2014}, decoupling the axial and transverse flows and generalizing the results of \citet{griffiths2008} to the case of multiple holes. However, they considered a prescribed axially varying viscosity but did not explicitly model the temperature. Rather they showed that the harmonic mean of the viscosity through the neck-down region determines the final fibre geometry, with the temperature to achieve this obtained by pulling with the required tension. The effect of adding pressure within air channels through the fibre was modelled by \citet{chen2015}. The recent work of \citet{JFM2019} coupled a temperature model to the flow model for drawing non-axisymmetric threads with arbitrary geometry. 
%The temperature model includes both heating in a furnace region and a cooling region that is appropriate for a typical fibre drawing process. 
The temperature model considers only advection and neglects conduction.
%Detailed reviews on the literature of fibre drawing process can be found in \cite{JFM2014, JFM2019}. 
%Heat conduction has not been included in any asymptotic model of fibre fabrication processes to date, whether extrusion or fibre drawing. Furthermore, temperature modelling (neglecting heat conduction) has only been done for fibre drawing, not extrusion. The aim of this paper is to precisely fill these gaps of modelling temperature with conduction in pulled extrusion and address the theoretical and numerical challenges that this entails. 
%The work in this paper is based on the model developed by \citet{JFM2019}. With pulling force dominating the gravitational force on the preform, the pulled extrusion is more similar to the fibre drawing model than the gravitational extrusion of \citet{JFM2017}.
%and \cite{JFM2017}, that includes the effects of axial conduction that are appropriate the extrusion process. This is done instead of having a model based on \citet{JFM2017} due to the fast speed of the pulled extrusion, where the pulling force dominates the weight of the glass, which is more similar to the fibre drawing model of \citet{JFM2019} than the gravitational extrusion of \citet{JFM2017}. 
%\citet{JFM2019} considered a pulled extrusion process for arbitrary cross-sectional geometry that they modelled as a steady-state problem with coupled fluid flow and a heat equation that neglects axial conduction.

This paper considers an arbitrary cross-sectional geometry but unlike all the above-mentioned studies includes the effects of axial conduction that are important in extrusion processes. In fibre drawing, both applied heating and cooling are important \citep{JFM2019}, but for preform extrusion the glass is hot when it exits the die and then cools down below it. Hence, we focus on external cooling. 

Having derived the model, we explain why it is extremely challenging to solve using standard numerical techniques. We hence develop a bespoke numerical method that is highly robust. We use this to consider the axisymmetric case and show that conduction can have a significant effect on the hole size at the exit of the device. 
\section{Model formulation}
%\subsection{Newtonian Incompressible Fluid}
\subsection{Full model}
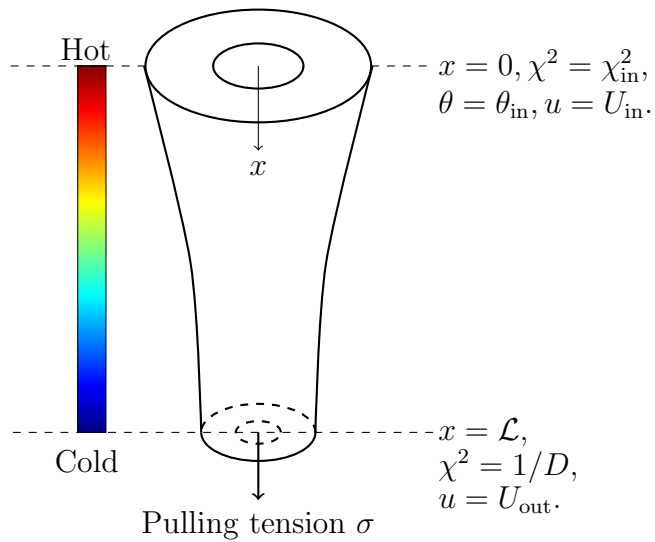
\begin{figure}
    \centering
    \begin{tikzpicture}[scale=0.75]
        %left curve
        \draw [thick] plot [smooth] coordinates {(-2.025,2.5) (-1.2,-1) (-1.0125,-4)};
        %right curve
        \draw [thick] plot [smooth] coordinates {(2.025,2.5) (1.2,-1) (1.0125,-4)};
        %outer tube
        \draw [thick] (0,2.5) ellipse (2 and 1);
        %inner tube
        \draw [thick] (0,2.5) ellipse (0.8 and 0.4);
  
        \begin{axis}[
            hide axis,
            scale only axis,
            height=0pt,
            width=0pt,
            xshift=-3.5cm,
            yshift=2.5cm,
            colormap/jet,
            colorbar horizontal,
            point meta min=18,
            point meta max=48,
            colorbar style={
                width=6.5cm,
                rotate=90,
                xtick={\empty},
                }
            ]
            \addplot [draw=none] coordinates {(0,0) (1,1)};
        \end{axis}

        % labels
        %explanation at x = 0
        \draw (3 ,2.5) node [anchor=west, text width=3.5cm]{$x=0,\chi^2 = \chi^2_\text{in}$,};
        \draw (3,1.8) node [anchor=west, text width=3.5cm]{$\theta = \theta_\text{in}, u = U_\text{in}.$};
        %top of heater
        \draw (-3.7,2.8) node [anchor=west, text width=3.5cm]{Hot};
 
        %bottom of heater
        \draw (-3.8,-4.5) node [anchor=west, text width=3.5cm]{Cold};
        \draw (3 ,-4) node [anchor=west, text width=3.5cm]{$x = \mathcal L$,};
        \draw (3 ,-4.6) node [anchor=west, text width=3.5cm]{$\chi^2 = 1/D$,};
        \draw (3 ,-5.2) node [anchor=west, text width=3.5cm]{$u = U_\text{out}$.};

        %description
        \draw[->,thick] (0,-4) -- (0,-5.2) node [anchor=north]{Pulling tension $\sigma$};
        \draw[->] (0,2.5) -- (0,1) node [anchor=north]{$x$};
        %\draw[<->] (-4.3,2.5) -- (-4.3,-4) node [rotate=90,midway,fill=white]{Deformation zone};
%        \draw[->] (5,0) -- (5,-2) node [anchor=north]{Gravity};
        
        % dashed curves on description at x = 0
        \draw [dashed] (-2,2.5) -- (-4.4,2.5);
        \draw [dashed] (2,2.5) -- (3.1,2.5);
        %bottom dashed line at x = L
        \draw [dashed] (3.1,-4) -- (-4.4,-4);
        %bottom solid line
        %\draw [thick] (-1.0125,-4) -- (1.0125,-4);

        \draw[thick,dashed] (1.0125,-4) arc(0:180:1.0125 and 0.50625);
        \draw[thick] (1.0125,-4) arc(0:-180:1.0125 and 0.50625); 
        \draw [thick,dashed] (0,-4) ellipse (0.4 and 0.2);
        \end{tikzpicture}
    \caption{Schematic diagram of extrusion with pulling tension $\sigma$.}
    \label{Fig:extrusion_diagram}
\end{figure}

A schematic diagram of the pulled extrusion process is shown in Figure \ref{Fig:extrusion_diagram}. The spatial coordinates are $(x,y,z)$, with the $x$ axis directed along the axis of the preform from the die exit at $x=0$, $t$ is time, the velocity vector is $\bm{u}=(u,v,w)$, pressure is $p$, and temperature is $\theta$. We define $\chi(x,t)$ to be the square root of the cross-sectional area at position $x$ and time $t$. The preform exits the die and enters the flow domain with cross-sectional area $\chi^2(0,t) = \chi^2_\text{in}$, %$\chi(0,t)$ being the square root of the cross-sectional area, 
temperature $\theta(0,y,z,t) = \theta_\text{in}$, axial speed $u(0,y,z,t) = U_\text{in}$, and a specified cross-sectional geometry. Below the die the preform cools and also stretches because it is pulled at a point sufficiently far from the die exit at a speed $U_\text{out} > U_\text{in}$. The ratio $U_\text{out}/U_\text{in}=D$ is known as the ``draw ratio''. We assume that at axial position $x=\mathcal L$ the preform is sufficiently cool, and the viscosity sufficiently large, that there is no significant deformation beyond this point. Thus the problem is quasi-steady and the evolution of the geometry can be modelled over the spatial domain $0 \leq x \leq \mathcal L$, with %a fixed position $x = L$ so that 
$u(\mathcal L,y,z,t)=U_\text{out}$.

As done by \citet{JFM2014,JFM2019,JFM2017}, and \citet{hayden_thesis}, we assume the glass to be an incompressible Newtonian fluid modelled by the continuity and %steady 
Navier-Stokes equations
\begin{subequations}
    \begin{align}
        \bm{\nabla \cdot u} &= 0, \label{Eq:Continuity}\\
        \rho \left(\frac{\partial\bm{u}}{\partial t} + \bm{u \cdot \nabla u}\right) &= - \bm{\nabla} p + \bm{\nabla \cdot \sigma} , \label{Eq:Newtonian}
    \end{align}
    \label{Eq:Newtonian_full}%
\end{subequations} 
where the fluid has %velocity $\bm{u} = (u,v,w)$, 
constant density $\rho$, %pressure $p$, temperature $\theta$, 
temperature-dependent viscosity $\mu(\theta)$, and $\bm{\sigma} = \mu(\theta) \left(\bm{\nabla u} + (\bm{\nabla u})^T \right)$ is the stress tensor.
We denote the location of the outer boundary by  $G^{(0)} (x,y,z,t) = 0$ and the location of the boundaries of the $N$ inner holes by  $G^{(i)} (x,y,z,t) = 0$ for $i=1,2,\ldots,N$. Then, the dynamic and kinematic boundary conditions %for $i = 1,2,...,N$ 
are
\begin{subequations}
    \begin{align}
        \bm{\sigma \cdot n}^{(i)} = -\gamma \kappa^{(i)} \bm{n} ^{(i)}, \label{Eq:dynamic_BC} \\
        \frac{\partial G^{(i)}}{\partial t} + \bm{u \cdot \nabla} G^{(i)} = 0, \label{Eq:kinematic_BC}
    \end{align}
    \label{Eq:kinematic_dynamic_BC}%
\end{subequations}
where $\gamma$ is the (constant) surface tension, $ \kappa^{(i)}$ is the local curvature of boundary $i$, and
$$\bm{n} ^{(i)} = \frac{\nabla G^{(i)}}{\left| \nabla G^{(i)} \right|}$$
is the outwards pointing normal vector on $G^{(i)} (x,y,z,t) = 0$. %for $i = 0,1,..., N$, where 
We denote the entire boundary by $G(x,y,z,t) = G^{(0)} + G^{(1)} + ... + G^{(N)}$. % denote the entire boundary. 
We also define $\Gamma(x,t)=\Gamma^{(0)}+\Gamma^{(1)}+\ldots +\Gamma^{(N)}$ to be the total boundary length of the cross-section at position $x$ and time $t$, where $\Gamma^{(0)}(x,t)$ is the length of the external cross-sectional boundary and $\Gamma^{(i)}(x,t)$ is the boundary length of the $i$th hole in the cross-section. %, and $\Gamma^{(0)}(x,t)$ to be external boundary length. \\

Note that the viscosity of glass is strongly dependent on temperature. However, surface tension is only weakly dependent on temperature \citep{Babcock, Shartsis, Parikh}, and so, is here assumed to be constant. In this paper we also assume zero air pressure in the internal channels since to our knowledge pressurization of air channels is not used in extrusion of fibre preform.

The temperature in the glass is modelled by
 \begin{equation}
     \rho c_p \left(\frac{\partial \theta}{\partial t} + \bm{u \cdot \nabla} \theta \right) = k \nabla^2 \theta, \label{Eq:heating}
 \end{equation}
 where $c_p$ is the specific heat capacity and $k$ is the conductivity of the glass, both assumed to be constant.
 On the external boundary we assume both radiative and convective (Newton) cooling, while on internal boundaries we assume no heat flux, giving the boundary conditions 
 \begin{subequations}
     \begin{align}
        -k \bm{\nabla} \theta \bm{\cdot n}^{(0)} &= k_b \beta \left(\theta^4 - \theta^4_a \right) + h_w \left(\theta-\theta_a\right) \text{ on $G^{(0)} = 0$,}  \label{Eq:external_BC_temp} \\
        -k \bm{\nabla} \theta \bm{\cdot n}^{(i)} &= 0 \text{ on $G^{(i)} = 0$, $i = 1,..., N$}. \label{Eq:internal_BC_temp}%
     \end{align}
 \end{subequations}
 The parameter $k_b$ is the Stefan--Boltzmann constant, $\beta$ is the surface absorptivity/emissivity, and $h_w$ is the convective heat transfer coefficient. The initial temperature of the preform is denoted by $\theta_\text{in}$, which is the temperature at $x = 0$. The (constant) ambient air temperature is denoted by $\theta_a$. At large distance from the inlet, the glass temperature must equilibriate with the ambient and hence
 \begin{equation*}
    \theta|_{x \rightarrow \infty} = \theta_a.
\end{equation*}

Typical values for all physical parameters are given in Table \ref{tab:parameters_val}. Note, however, that the values of $\beta$ and $h_w$ differ from one device to another and can be somewhat challenging to estimate. 

 \begin{table}
    \centering
    \begin{tabular}{c|c|c|c}
        Parameter & Symbol & Approx. value & Units \\
        \hline
        Specific heat & $c_p$ & 557 & $\text{J kg}^{-1} \text{ K} ^{-1}$ \\
         Conductivity & $k$ & 0.78 & $\text{W m}^{-1} \text{ K}^{-1}$ \\
         Density & $\rho$ & 3600 & $\text{kg m}^{-3}$ \\
         Stefan--Boltzmann constant & $k_b$ & $5.67 \times 10^{-8}$ & $\text{W m}^{-2} \text{ K}^{-4}$ \\
         Surface absorptivity/emissivity & $\beta$ & 0.8 & - \\
         Convective heat transfer coefficient & $h_w$ & 95 & $\text{W m}^{-2} \text{ K}^{-1}$ \\
         Ambient air temperature & $\theta_a$ & 290 & \text{K} \\
         Surface tension & $\gamma$ & 0.24 & $\text{N m}^{-1}$ \\
          Initial temperature & $\theta_\text{in}$ & 970 & \text{K} \\
         Viscosity at temperature $\theta_\text{in}$ & $\mu_\text{in}$ & $2 \times 10^{4}$ & Pa s \\
         Preform cross-sectional area & $\chi_\text{in}^2$ & $10^{-4}$ & $\text{m}^2$ \\
         Draw ratio & $D$ & $\mathcal{O}(1)$ & - \\
         Length & $\mathcal{L}$ & 0.2 & m \\
         Feed speed & $U_\text{in}$ & $10^{-5}$ & $\text{m s}^{-1}$
    \end{tabular}
    \caption{Typical parameter values for pulled extrusion based on F2 glass \citep{JFM2019,hayden_thesis,Schott_data}.}
    \label{tab:parameters_val}
\end{table}

In addition we require a viscosity--temperature function $\mu(\theta)$, which, in this paper, is assumed to be of the form used by \citet{he2016extension},
\begin{equation}
    \mu(\theta) = \mu_\text{in}e^{-B \left(\theta - \theta_\text{in}\right)},
    \label{Eq:viscosity_old}
\end{equation}
where $\mu_\text{in}$ is the viscosity at the initial temperature $\theta_\text{in}$, and $B$ is the constant that governs the sensitivity of the viscosity to temperature. This exponential law models the dramatic changes that occur in viscosity during the cooling process.

\subsection{Non-dimensionalisation}\label{sec:dimensionlessmodel}
We non-dimensionalise the equations using a similar approach to that of \citet{JFM2019}. The axial position $x$ is non-dimensionalised by the typical length $\mathcal{L}$ over which the preform deforms, whereas the transverse variables $(y,z)$ are non-dimensionalised using $\chi_\text{in}$ that represents the square root of the initial cross-sectional area. We define the slenderness parameter $\varepsilon$ to be the ratio of the transverse length scale to the axial length scale,
\begin{equation}
    \varepsilon = \frac{\chi_\text{in}}{\mathcal{L}},
    \label{Eq:epsilon}
\end{equation}
and assume a slender preform, so that $\varepsilon \ll 1$. The dimensionless variables, denoted by primes are, 
\begin{subequations}
    \begin{equation*}
       (x,y,z) = \mathcal{L}(x', \varepsilon y', \varepsilon z'), \quad t = \frac{\mathcal{L}}{U_\text{in}}t', \quad (u,v,w) = U_{in}(u', \varepsilon v', \varepsilon w'), \\ 
    \end{equation*}
    \begin{equation*}
        \chi = \chi_{in} \chi', \quad \Gamma = \chi_\text{in} \Gamma', \quad \kappa = \frac{\kappa '}{\chi_\text{in}}, \quad \theta = \theta_\text{in}\theta', \quad \mu = \muin \mu', \quad p = \frac{\muin U_\text{in}}{\mathcal{L}}p'.
    \end{equation*}
\end{subequations}
%where primes denote dimensionless variables.\\
%
%Time is non-dimensionalised using the time needed to extrude a typical preform, obtained by dividing the preform length scale $\mathcal{L}$ by the feed speed $U_\text{in}$. The feed speed $U_\text{in}$ is used to non-dimensionalise the velocity $u$ in the $x$ direction while the velocities $v$ and $w$ in the cross-sectional plane are scaled using $\varepsilon U_\text{in}$. The square root of the cross-sectional area $\chi$, temperature $\theta$, and the viscosity $\mu(\theta)$ are all non-dimensionalised by their respective initial values, whereas the boundary length $\Gamma$ and curvature $\kappa$ are non-dimensionalised by the root of initial cross-sectional area. Since the preform starts hot at $\theta' = 1$ and it cools down throughout the process, the value of $\theta'$ for $x>0$ will be less than one.\\
%
The dimensionless form of the viscosity--temperature function \eqref{Eq:viscosity_old} is 
\begin{equation}
    \mu'(\theta') = e^{C(1-\theta')},
    \label{Eq:viscosity_new}
\end{equation}
where $C = B \theta_\text{in}$.

%The dimensionless parameter $C$ governs the sensitivity of the viscosity to temperature and depends on the glass and the temperature range in the extrusion process. %% this is moved to results section\\

% The non-dimensionalised outwards pointing normal vector is given by
% $$\bm{n}^{'(i)} = \frac{1}{\left| \nabla' G^{(i)} \right|} \left(\varepsilon \frac{\partial G^{(i)}}{\partial x'}, \frac{\partial G^{(i)}}{\partial y'}, \frac{\partial G^{(i)}}{\partial z'}\right),$$
% from which we define,
% $$ n_{x'}^{'(i)} = \frac{1}{\left| \nabla' G^{(i)} \right|} \frac{\partial G^{(i)}}{\partial x'}, \qquad  \bm{n}^{'(i)}_{\perp'} = \frac{1}{\left| \nabla G^{(i)} \right|} \left(\frac{\partial G^{(i)}}{\partial y'}, \frac{\partial G^{(i)}}{\partial z'}\right) = \left(n_{y'}^{'(i)}, n_{z'}^{'(i)} \right).$$ \\

% The non-dimensionalised continuity equation \eqref{Eq:Continuity} is 
%In a similar manner to \cite{JFM2019}, 

We now use these scalings to non-dimensionless our problem. Also, because it is quasi-steady, we assume time independence. The continuity equation becomes
\begin{equation}
    \frac{\partial u'}{\partial x'} + \frac{\partial v'}{\partial y'} + \frac{\partial w'}{\partial z'} = 0,
    \label{eq:nondim_continuity}
\end{equation}
while
%We proceed by considering the steady-state problem and hence drop the time derivatives and non-dimensionalising 
the momentum equation \eqref{Eq:Newtonian} yields
\begin{subequations}
\begin{align}
     &\varepsilon^2 \Rey \left(u' \frac{\partial u'}{\partial x'} +  v' \frac{\partial u'}{\partial y'} + w' \frac{\partial u'}{\partial z'} \right)  = -\varepsilon^2 \frac{\partial p'}{\partial x'} + \varepsilon^2 \frac{\partial}{\partial x'} \left(2\mu' \frac{\partial u'}{\partial x'}\right) \nonumber\\
    &\quad + \frac{\partial}{\partial y'} \left(\mu' \left[\frac{\partial u'}{\partial y'} + \varepsilon^2\frac{\partial v'}{\partial x'}\right]\right) + \frac{\partial}{\partial z'} \left(\mu' \left[\frac{\partial u'}{\partial z'} + \varepsilon^2\frac{\partial w'}{\partial x'}\right]\right),
    \label{eq:nondim_momentum}\\
% \end{align}
%
% The $y$ and $z$ directions for the momentum equation  \eqref{Eq:Newtonian} yield 
% \begin{align}
    &\varepsilon^2 \Rey \left(u' \frac{\partial v'}{\partial x'} + \varepsilon v' \frac{\partial v'}{\partial y'} + \varepsilon w' \frac{\partial v'}{\partial z'} \right) =-\frac{\partial p'}{\partial y'}
    + \frac{\partial}{\partial x'}\left(\mu' \left[ \varepsilon^2 \frac{\partial v'}{\partial x'} + \frac{\partial u'}{\partial y'}\right]\right) \nonumber\\
    &\quad + \frac{\partial}{\partial y'} \left(2\mu' \frac{\partial v'}{\partial y'}\right)  + \frac{\partial}{\partial z'} \left(\mu' \left[\frac{\partial v'}{\partial z'} + \frac{\partial w'}{\partial y'}\right]\right),
    \label{eq:nondim_momentum_y}\\
% \end{align} 
% and
% \begin{align}
    &\varepsilon^2 \Rey \left(u' \frac{\partial w'}{\partial x'} + \varepsilon v' \frac{\partial w'}{\partial y'} + \varepsilon w' \frac{\partial w'}{\partial z'} \right) =-\frac{\partial p'}{\partial z'}
    + \frac{\partial}{\partial x'}\left(\mu' \left[\varepsilon^2 \frac{\partial w'}{\partial x'} + \frac{\partial u'}{\partial z'}\right]\right) \nonumber \\
    &\quad + \frac{\partial}{\partial y'} \left(\mu' \left[\frac{\partial w'}{\partial y'} + \frac{\partial v'}{\partial z'}\right]\right)  + \frac{\partial}{\partial z'}\left(2\mu' \frac{\partial w'}{\partial z'}\right),
    \label{eq:nondim_momentum_z}
\end{align}
\end{subequations}
where 
\begin{equation}
\Rey = \frac{\rho U_\text{in} \mathcal{L}}{\mu_\text{in}}
\label{eq:reynolds}
\end{equation}
is the Reynolds number.
Similar non-dimensionalisation of the dynamic boundary condition \eqref{Eq:dynamic_BC} gives 
\begin{subequations}
    \begin{align}
        &2\mu' \varepsilon^2 \frac{\partial u'}{\partial x'} n_{x'}^{'(i)} + \mu' \left(\frac{\partial u'}{\partial y'}+\varepsilon^2 \frac{\partial v'}{\partial x'}\right)n_{y'}^{'(i)}+ \mu' \left(\frac{\partial u'}{\partial z'}+\varepsilon^2 \frac{\partial w'}{\partial x'}\right)n_{z'}^{'(i)}\nonumber\\
        &\quad = -\varepsilon^2 \frac{\gamma \kappa'^{(i)}}{\Ca} n_{x'}^{'(i)},\label{axialBC} \\
        &2\mu' \frac{\partial v'}{\partial y'} n_{y'}^{'(i)} + \mu' \left(\varepsilon^2 \frac{\partial v'}{\partial x'}+ \frac{\partial u'}{\partial y'}\right)n_{x'}^{'(i)}+ \mu' \left(\frac{\partial v'}{\partial z'}+\varepsilon^2 \frac{\partial w'}{\partial y'}\right)n_{z'}^{'(i)}\nonumber \\
        &\quad = -\frac{\gamma \kappa'^{(i)}}{\Ca} n_{y'}^{'(i)}, \\
        &2\mu' \frac{\partial w'}{\partial z'} n_{z'}^{'(i)} + \mu'  \left(\varepsilon^2 \frac{\partial w'}{\partial x'}+ \frac{\partial u'}{\partial z'}\right)n_{x'}^{'(i)}+ \mu' \left(\frac{\partial v'}{\partial z'}+\varepsilon^2 \frac{\partial w'}{\partial y'}\right)n_{y'}^{'(i)}\nonumber\\
        &\quad = -\frac{\gamma \kappa'^{(i)}}{\Ca} n_{z'}^{'(i)},
    \end{align}
    \label{eq:nondim_dynamic_BC}%
\end{subequations}
where 
\begin{equation}
    \Ca = \frac{\mu_\text{in} U_\text{in} \chi_\text{in}}{\gamma \mathcal{L}}
    \label{Eq:ca}
\end{equation} 
is the capillary number, and the kinematic boundary conditions \eqref{Eq:kinematic_BC} become
\begin{equation}
      u' \frac{\partial G^{(i)}}{\partial x'} + v' \frac{\partial G^{(i)}}{\partial y'} + w' \frac{\partial G^{(i)}}{\partial z'} = 0, \quad i = 0,1,...,N.
     \label{eq:nondim_kinematic_BC}
\end{equation}
We also have the boundary conditions $u'=1$ at $x'=0$ and $u'=D$ at $x'=1$.

Finally, non-dimensionalising the temperature equation \eqref{Eq:heating} and assuming no time dependence %dropping the time dependent term 
gives
\begin{equation}
    \varepsilon^2 \Pe \left(u' \frac{\partial \theta'}{\partial x'} + v' \frac{\partial \theta'}{\partial y'} + w' \frac{\partial \theta'}{\partial z'}  \right) = \varepsilon^2 \frac{\partial^2 \theta'}{\partial x'^2} + \frac{\partial^2 \theta'}{\partial y'^2} + \frac{\partial^2 \theta'}{\partial z'^2},
    \label{eq:nondim_temp}
\end{equation}
where
\begin{equation}
    \Pe = \frac{\rho c_p U_{in} \mathcal{L}}{k}
    \label{Eq:Peclet}
\end{equation}
is the P\'eclet number.
%On non-dimensionalising 
The boundary conditions \eqref{Eq:external_BC_temp} and \eqref{Eq:internal_BC_temp} become %we find
\begin{subequations}
\begin{align}
    &-\left(\varepsilon^2 \frac{\partial \theta'}{\partial x'} n_{x'}^{'(0)} + \bm{\nablap}' \theta \bm{\cdot n}^{'(0)}_{\perp'} \right) = \varepsilon^2 \Pe \left[\HR \left(\theta'^4 - \theta'^4_a \right) + \HC \left(\theta'-\theta'_a\right) \right],
    \label{eq:nondim_ext_BC_temp}\\
    &-\left(\varepsilon^2 \frac{\partial \theta'}{\partial x'} n_{x'}^{'(i)} + \bm{\nablap}' \theta \bm{\cdot n}^{'(i)}_{\perp'} \right) = 0\quad \text{for}\ i = 1,..., N.
    \label{eq:nondim_int_BC_temp}
\end{align}
\label{eq:nondim_BC_temp}
\end{subequations}
where
%\begin{subequations}
    \begin{align}
        \HR = \frac{\beta k_b \theta_\text{in}^3 \mathcal{L}}{\rho c_p U_\text{in} \chi_\text{in}}, %\label{eq:hr_def_new}
        \quad\HC = \frac{h_w \mathcal{L}}{\rho c_p U_\text{in} \chi_\text{in}}.
        \label{eq:hc_def_new}%
    \end{align}
%\end{subequations}
% and 
% $$
% \bm{\nablap}' = \left( \frac{\partial}{\partial y'}, \frac{\partial}{\partial z'} \right).
% $$

%Note that the parameters $\HR$ and $\HC$ are equivalent to $\HR/\Ca$ and $\HC/\Ca$ as defined by \cite{JFM2019}, adjusting also for our different scaling of $\theta$. 
% Non-dimensionalising the temperature boundary condition \eqref{Eq:internal_BC_temp} on the internal holes gives
% \begin{equation}
%     -\left(\varepsilon^2 \frac{\partial \theta'}{\partial x'} n_{x'}^{'(i)} + \bm{\nablap}' \theta \bm{\cdot n}^{'(i)}_{\perp'} \right) = 0 \text{ for $i = 1,..., N$}.
%     \label{eq:nondim_int_BC_temp}
% \end{equation} \\

%May need BCs, maybe not necessary
These equations are similar to those of \citet{JFM2019} but differ in that our problem has no initial heating region, just cooling below the die exit at $x'=0$ where the temperature is at its maximum value $\theta'=1$. As $x'\rightarrow\infty$ we have $\theta'\rightarrow \theta_a/\theta_\text{in}$. %from the maximum temperature at $x = 0$. % below which the glass cools. 
%In this our model is similar to those of %the model obtained by 
%\cite{hayden_thesis} and \cite{JFM2017}. %, where the glass cools down below $x = 0$. 

\subsection{Asymptotic model}
We next make use of the smallness of the slenderness parameter $\varepsilon\ll 1$ to obtain an asymptotic model. Details of the derivation process are similar to those given in \cite{JFM2019} and we here just give a brief outline and refer the reader to \cite{JFM2019} for details.

First, all dependent variables %$u'$, $v'$, $w'$, $\theta'$ 
are expanded in powers of $\varepsilon^2$ %, where
\begin{equation}
     u' = u_0(x',y',z') + \varepsilon^2 u_1(x',y',z') + \varepsilon^4 u_2 (x',y',z') + \hdots,  \label{eq:asymptotics} 
\end{equation}
% \begin{subequations}
%     \begin{align}
%         u' &= u_0(x,y,z) + \varepsilon^2 u_1(x,y,z) + \varepsilon^4 u_2 (x,y,z) + \hdots  \label{eq:asymptotics_u} \\
%         % v' &= v_0(x,y,z) + \varepsilon^2 v_1(x,y,z) + \varepsilon^4 v_2 (x,y,z) + \hdots  \label{eq:asymptotics_v} \\
%         % w' &= w_0(x,y,z) + \varepsilon^2 w_1(x,y,z) + \varepsilon^4 w_2 (x,y,z) + \hdots  \label{eq:asymptotics_w} \\
%         % \theta' &= \theta_0(x,y,z) + \varepsilon^2 \theta_1(x,y,z) + \varepsilon^4 \theta_2 (x,y,z) + \hdots \label{eq:asymptotics_theta}
%     \end{align}
%     \label{eq:asymptotics}%
% \end{subequations}
with similar expansions for $v'$, $w'$, $\theta'$, $p'$,  $\chi'$, $\Gamma'^{(i)}$, $\kappa'^{(i)}$, and $G^{(i)}$. These asymptotic expansions are substituted into the equations of section~\ref{sec:dimensionlessmodel}
%(\ref{eq:nondim_continuity}, ~\ref{eq:nondim_momentum}, ~\ref{eq:nondim_momentum_y}, ~\ref{eq:nondim_momentum_z}, ~\ref{eq:nondim_dynamic_BC}, ~\ref{eq:nondim_kinematic_BC}, ~\ref{eq:nondim_temp}, ~\ref{eq:nondim_ext_BC_temp}, ~\ref{eq:nondim_int_BC_temp}),
and equations at each order of $\varepsilon$ are obtained. %leading-order equations are obtained.
%In this section, the substitution of the asymptotic expansion of the temperature $\theta'$ will be shown to illustrate the process. \\

% Substituting \eqref{eq:asymptotics} into equation \eqref{eq:nondim_temp}, and taking the leading order terms we obtain
% \begin{equation}
%      \nablap^2 \theta_0 = 0.
%      \label{eq:Laplacian}
% \end{equation} \\

% Similarly the leading-order non-dimensionalised external boundary condition \eqref{eq:nondim_ext_BC_temp} is given by
% \begin{equation}
%     \bm{\nablap} \theta_0 \bm{\cdot n}^{(0)}_\perp = 0
%     \label{eq:perp_zero}
% \end{equation}
% and the leading-order internal boundary conditions \eqref{eq:nondim_int_BC_temp} are
% \begin{equation}
%     \nablap \theta_0 \bm{\cdot n}^{(i)}_\perp = 0, \text{ for } i = 1, 2, \hdots, N. 
%     \label{eq:perp_i}
% \end{equation}
%Equations \eqref{eq:Laplacian}, \eqref{eq:perp_zero} and \eqref{eq:perp_i} imply that
%\begin{equation}
 %   \frac{\partial \theta_0}{\partial y} = 0, \quad \frac{\partial \theta_0}{\partial z} = 0. 
%\end{equation}
% We know that $\theta$ is independent of $y$ and $z$, and so we write
% \begin{equation}
%     \theta_0 = \theta_0(x).
%     \label{eq:theta_depends_x}
% \end{equation} 
% Since the viscosity is a function of temperature $\mu(\theta)$, the viscosity is also independent of $y$ and $z$, and assuming $\varepsilon^2 \Rey \ll 1$ and following a procedure similar to \cite{JFM2019}, we obtain
% \begin{equation}
%     u_0 = u_0(x).
%     \label{eq:u_depends_x}
% \end{equation} 
It is readily shown \citep{JFM2019} that the leading-order temperature $\theta_0$ and velocity $u_0$ are independent of the cross-sectional coordinates, i.e. 
$$
\theta_0 = \theta_0(x'), \qquad u_0 = u_0(x').
$$
Furthermore, the leading-order conservation equation integrated over the cross-section at axial position $x'$, together with the leading-order
kinematic condition integrated over the cross-section boundary and a transport theorem from \citet{Dew_transport}, give  %the conservation of mass, where
\begin{equation}
    u_0 \chi_0^2 = 1,
    \label{eq:chi_u_steady}
\end{equation}
for any axial position $x'$. 
% Next, we will derive the leading-order temperature equation. We take the $\mathcal{O}(\varepsilon^2)$ terms from the asymptotic expansion of \eqref{eq:nondim_temp} and, using the fact that $\theta_0(x,t)$ is independent of $y$ and $z$, we obtain
% \begin{equation}
%     \Pe u_0 \frac{\partial \theta_0}{\partial x} - \frac{\partial^2 \theta_0}{\partial x^2} = \nablap^2 \theta_1.
%     \label{eq:asymptotics_theta_epsilon2_rearranged}
% \end{equation}
% The $\mathcal{O}(\varepsilon^2)$ terms are also taken from the asymptotic expansions of the external and internal boundary conditions \eqref{eq:nondim_ext_BC_temp} and \eqref{eq:nondim_int_BC_temp}, giving
% \begin{equation}
%     -\left(\frac{\partial \theta_0}{\partial x} n_{x}^{(0)} + \bm{\nablap} \theta_1 \bm{\cdot n}^{(0)}_\perp \right) = \Pe \left[\HR \left(\theta_0^4 - \theta^4_a \right) + \HC \left(\theta_0-\theta_a\right) \right] \text{ on $G^{(0)} = 0$,} \label{eq:asymptotics_int}
% \end{equation}
% and 
% \begin{equation}
%     -\left(\frac{\partial \theta_0}{\partial x} n_{x}^{(i)} + \bm{\nablap} \theta_1 \bm{\cdot n}^{(i)}_\perp \right) = 0 \text{ on $G^{(i)} = 0$, $i = 1,..., N$}, \label{eq:asymptotics_ext}
% \end{equation}
% respectively.\\
% After integration of \eqref{eq:asymptotics_theta_epsilon2_rearranged} over the cross-section and using a transport theorem derived by \cite{Dew_transport}, we obtain
%Following a similar asymptotic procedure to that used by \cite{JFM2019}

Then, considering the $O(\varepsilon^2)$ terms of the momentum equation \eqref{eq:nondim_momentum}, integrating over the cross-section at axial position $x'$, and using the divergence theorem, the $O(\varepsilon^2)$ terms of boundary condition \eqref{axialBC}, a transport theorem given by  \citet{Dew_transport}, and the conservation equation \eqref{eq:chi_u_steady}, we obtain the leading-order axial-flow equation
\begin{equation}
    \Rey \frac{d u_0}{d x'} = \frac{d}{d x'}\left(3\mu(\theta_0) \chi^2_0 \frac{d u_0}{d x'}\right) + \frac{1}{2\Ca}\frac{d  \Gamma_0}{d x'},
    \label{eq:axial_flow}
\end{equation}
where $\Gamma_0(x')$ is the total leading-order boundary length of the cross-section at position $x'$.
This may be integrated with respect to $x'$ and further manipulated to yield
\begin{equation}
    \frac{\Rey}{\chi_0^2} + \frac{6\mu(\theta_0)}{\chi_0} \frac{d \chi_0}{d x'} - \frac{\Gamma_0}{2\Ca} = -6\sigma,
    \label{eq:axial_flow2}
\end{equation}
where $6\sigma$ is the pulling tension.
%Similarly, considering the $\mathcal{O}(\varepsilon^2)$ terms of \eqref{eq:nondim_temp} and \eqref{eq:nondim_BC_temp}, integrating over the cross-sectional area and using the boundary conditions \eqref{eq:nondim_ext_BC_temp} and \eqref{eq:nondim_int_BC_temp} and a transport theorem, % given by \cite{Dew_transport},

In a similar manner
we obtain the leading order temperature equation
\begin{align}
      \frac{d \theta_0}{d x'} =\frac{1}{\Pe}\frac{d}{d x'}\left(\chi_0^2 \frac{d \theta_0}{d x'} \right) -\Gamma_0^{(0)} \left[\HR \left(\theta_0^4 - \theta^4_a \right)+ \HC \left(\theta_0-\theta_a\right) \right].
     \label{eq:temperature}
\end{align}
Here $\Gamma^{(0)}(x')$ is the leading-order length of the external boundary of the cross-section at axial position $x'$. 
%
% Performing similar steps on the leading order kinematic condition from equation \eqref{eq:nondim_kinematic_BC} and integrating over the boundary gives 
% \begin{equation}
%     \frac{\partial}{\partial x}\left(u_0\chi_0^2\right) = 0,
%     \label{eq:conservation_mass_u_chi}
% \end{equation}
% which represents the conservation of mass. After integration and applying the boundary conditions we obtain
% Finally, the leading-order conservation equation and
% kinematic condition gives %the conservation of mass, where
% \begin{equation}
%     u_0 \chi_0^2 = 1.
%     \label{eq:chi_u_steady}
% \end{equation} 
%This represents a constant value of the product of speed and cross-sectional area throughout the length of the glass. 

The above equations must be solved subject to boundary conditions $\chi_0(0)=1$, $\theta_0(0)=1$, and $\theta_0(\infty)=\theta_a/\theta_\text{in}$. The tension parameter $\sigma$ must be chosen to satisfy $\chi^2_0(1)=1/D$. %From \eqref{eq:chi_u_steady} we readily obtain 
%$\chi_0^2(1) = 1/D$. 

The leading-order boundary lengths $\Gamma_0(x')$ and $\Gamma_0^{(0)}(x')$, appearing in \eqref{eq:axial_flow}, \eqref{eq:axial_flow2} and \eqref{eq:temperature}, must be obtained from a transverse flow model describing how the cross-sectional geometry changes with $x'$. As described in \cite{JFM2014,JFM2019}, using appropriate scalings and variable transformations, this can be written as a classical 2D Stokes-flow problem with unit coefficient of surface tension in a domain of unit area which gives the geometry evolution from some arbitrarily shaped initial geometry at $x'=0$ to the final geometry at $x'=1$. However, for simplicity, in this paper we consider extrusion of an axisymmetric tube having an annular cross-section.

%\subsection{Leading-order transverse-flow model}
%The system is significantly simpler if we consider an axisymmetric glass tube having an annular cross-section. 
We define $R_0(x')$ to be the (leading-order) outer radius of the preform and $\phi_0(x')$ to be the (leading-order) ratio of the internal radius to the external radius, so the internal radius is defined as $\phi_0 R_0$. Therefore, the cross-sectional area is given by
\begin{equation}
    \chi_0^2 = \pi R_0^2(1 - \phi_0^2).
    \label{eq:chi_R}
\end{equation}
Although the deformation can be described in terms of $\chi_0^2$ and $\phi_0$ we choose to follow \citet{JFM2014,JFM2019}, and describe it in terms of $\chi_0^2$ and $\alpha_0$, where $\chi_0\alpha_0=R_0(1-\phi_0)$ is the (leading-order) wall thickness of the cross-section. Then, using \eqref{eq:chi_R}, we have
\begin{equation}
    \alpha_0 = \sqrt{\frac{\left(1-\phi_0\right)}{\pi(1+\phi_0)}},
    \label{Eq:alpha_phi}
\end{equation}
the wall thickness scaled with $\chi_0$ (or the wall thickness of an annulus with unit cross-sectional area).
The scaled wall thickness $\alpha_0$ is constrained by the condition $0 < \alpha_0 \le 1/\sqrt{\pi}$. The value $\alpha_0 = 0$ corresponds to the wall bursting ($\phi_0\rightarrow 1$) and is the minimum value of $\alpha_0$, whereas the maximum value $\alpha_0 = 1/\sqrt{\pi}$ corresponds to the hole closing ($\phi_0=0$). It is readily shown that %We also utilise the definition of 
the length of the external cross-sectional boundary is 
\begin{equation}
    \Gamma_0^{(0)} = \chi_0\left(\frac{1}{\alpha_0} + \pi \alpha_0 \right)
    \label{eq:gamma0}
\end{equation}
and the total cross-sectional boundary length is
\begin{equation}
    \Gamma_0 = \frac{2 \chi_0}{\alpha_0}.
    \label{eq:gamma}
\end{equation} 

The leading-order transverse-flow model in \cite{JFM2014,JFM2019} is derived using a transformation from the axial coordinate $x'$ to a Lagrangian-like variable identifying a fluid cross-section, and all differential equations are then written in terms of this variable. In this paper, however, because we are considering only extrusion of axisymmetric tubes, we have, for simplicity, chosen to retain the Eulerian coordinate $x'$, in terms of which it is readily shown that $\alpha_0$ is given by the differential equation %use that Lagrangian-like variable, but the leading-order transverse-flow model has been derived in terms of the independent variables $x,t$ by \cite{Fitt2002}. We here use the \cite{Fitt2002} work but write this in terms of the dependent variables $\chi,\alpha$. Taking equation (21) from \cite{Fitt2002} and write this in terms of $\chi_0^2, \alpha$, then using the conservation of mass \eqref{eq:chi_u_steady}, we obtain the transverse-flow equation 
\begin{equation}
     \frac{d \alpha_0}{d x'} = \frac{\chi_0}{2\, \Ca\, \mu(\theta_0)}.
    \label{eq:transverse_flow_zero}
\end{equation}

From this point on, the zero subscripts denoting the leading-order terms and the primes on dimensionless variables will, for convenience, be dropped.

\subsection{The model as a system of first-order ODEs}
%Using the conservation of mass equation \eqref{eq:chi_u_steady}, the axial velocity $u$ may be eliminated from our equations to yield 
Our model is a system of coupled ordinary differential equations (ODEs) for $\chi$, $\theta$, and $\alpha$, namely \eqref{eq:axial_flow2}, \eqref{eq:temperature}, and \eqref{eq:transverse_flow_zero}; each equation involves all three dependent variables. %Therefore, we must solve the leading-order steady state model as a system of ordinary differential equations.% as done by \citep{JFM2019}, shown in \eqref{Eq:second_order}. However, the model in this paper is with respect to $x$ instead of a Lagrangian-like variable as done by \cite{JFM2019}. Also, we have removed the time derivative and convert the partial derivatives to regular derivatives. \\
Using typical physical parameter values from Table~\ref{tab:parameters_val} we find that the Reynolds number \eqref{eq:reynolds} is very small,  $\Rey \equiv \mathcal{O}(10^{-8})$, which justifies neglect of the inertial term in \eqref{eq:axial_flow2}. %We also integrate this equation to obtain a first-order ordinary differential equation for $\chi$.
%Hence, neglecting inertia in \eqref{eq:axial_flow} while expressing it in terms of $\alpha$, also using \eqref{eq:chi_u_steady} and integrating with respect to $x$, we obtain \eqref{Eq:second_chi}. Similarly, using the conservation of mass from \eqref{eq:chi_u_steady} on \eqref{eq:transverse_flow_zero}, we obtain \eqref{Eq:second_alpha}.\\

%Using the conservation of mass \eqref{eq:chi_u_steady} and the temperature equation \eqref{eq:temperature}, expressing the equation in terms of $\alpha$, 
Previous work \citep{JFM2019, Taroni2013} on fibre drawing assumed a large P\'{e}clet number, enabling neglect of the second-order term in the temperature equation. Here, because we are concerned with the much slower process of preform extrusion, we assume $\Pe = \mathcal{O}(1)$ so that energy conduction, as well as advection, is important. Thus, we have a second-order ODE for $\theta$ coupled with two first-order ODEs for $\chi$ and $\alpha$. %The second-order conduction term is retained in this paper since it is important, assuming that $\Pe = \mathcal{O}(1)$, instead of large $\Pe$ as done by \cite{JFM2019}. 
%Since this temperature model is coupled to the first-order flow model, 
Hence, we write the second-order temperature equation as two coupled first-order ODEs.
%\eqref{Eq:second_theta} and \eqref{Eq:second_y}. 
To do this we define %This is done by defining $d\theta/dx$ differently from the standard $d\theta/dx = y$, 
$$
y = \frac{\chi^2}{\Pe} \frac{d \theta}{d x},
$$
which leads to the system of equations% to code and solve. \\
% Substituting the relationship \eqref{eq:chi_u_steady} into the axial-flow equation given in equation \eqref{eq:axial_flow}, while neglecting inertia and expressing the equation in terms of $\alpha$, we obtain
% \begin{equation*}
%    \frac{\partial}{\partial x}\left(-6\mu(\theta) \frac{1}{\chi} \frac{\partial \chi}{\partial x}\right) + \frac{1}{\Ca}\frac{\partial}{\partial x}\left(\frac{\chi}{\alpha}\right) = 0.
% \end{equation*}
% Integrating with respect to $x$ and rearranging we have
% \begin{equation}
%    \frac{\partial \chi}{\partial x}  = \frac{\chi^2}{6 \alpha \Ca \mu(\theta)} - \frac{\chi \sigma}{\mu(\theta)}.
%    \label{final_chi_steady}
% \end{equation}
% where $6\sigma$ is the dimensionless tension as defined by \cite{JFM2014, JFM2019}. Using the conservation of mass \eqref{eq:chi_u_steady} and the temperature equation \eqref{eq:temperature}, expressing the equation in terms of $\alpha$, we have 
% \begin{align}
%      \frac{\partial \theta}{\partial x} =\frac{1}{\Pe}\frac{\partial}{\partial x}\left(\chi^2 \frac{\partial \theta}{\partial x} \right) -\chi\left(\frac{1}{\alpha} + \pi \alpha \right) \left[\HR \left(\theta^4 - \theta^4_a \right)+ \HC \left(\theta-\theta_a\right) \right].
%      \label{eq:asymptotics_theta_steady}
% \end{align} 
% Similarly, using \eqref{eq:chi_u_steady} and \eqref{eq:transverse_flow_zero}, we obtain
% \begin{equation}
%      \frac{\partial \alpha}{\partial x}= \frac{\chi}{2 \Ca \mu(\theta)}.
%     \label{eq:alpha_steady}
% \end{equation} \\
\begin{subequations}
    \begin{align}
        \frac{d\chi}{dx} &= \frac{\chi^2}{6 \alpha \Ca \mu(\theta)} - \frac{\chi\sigma}{\mu(\theta)} \text{ , }\label{Eq:second_chi}\\
        \frac{d\theta}{dx} &= 
        \begin{dcases}
            \frac{\Pe}{\chi^2}y \text{ , } &\theta > \theta_a,\\
            0 \text{ , } &\theta \leq \theta_a,
        \end{dcases}
        \label{Eq:second_theta}\\
        \frac{dy}{dx} &= \frac{\Pe}{\chi^2}y + \chi \left(\frac{1}{\alpha} + \pi \alpha \right)\left[\HR \left(\theta^4 - \theta_a^4\right)+\HC \left(\theta - \theta_a\right) \right]  \text{ , } \label{Eq:second_y} \\
     \frac{d\alpha}{dx} &= 
        \begin{dcases}
            \frac{\chi}{2\Ca\mu(\theta)} \text{ ,} &\alpha < 1/\sqrt{\pi},\\
            0\text{ , } &\alpha \geq 1/\sqrt{\pi}.
        \end{dcases}
        \label{Eq:second_alpha}
    \end{align}
        \label{Eq:second_order}%
\end{subequations}

 We ensure that $\alpha$ does not exceed its maximum value of $1/\sqrt{\pi}$ by requiring its gradient to be zero once this value is reached. We assume the value of $\chi \ne 0$ since this would represent breaking of the thread, which is not a desirable extrusion outcome.
 %In a similar manner, we also ensure that the temperature not fall below its minimum value of ambient air temperature. 
 
 We need to solve this system of equations over the domain $0\le x\le 1$ where deformation occurs. The boundary conditions at $x= 0$ are given by
\begin{equation}
    \chi|_{x = 0} = 1, \qquad \qquad \theta|_{x = 0} = 1, \qquad \qquad \alpha|_{x = 0} = \alpha_\text{in}.
        \label{Eq:original_BC_zero}
\end{equation}
% \theta|_{x \rightarrow \infty} = \theta_a,\quad \quad
We also have the condition where $\theta \rightarrow \theta_a$ as $x \rightarrow \infty$. However, because 
our solution domain is, necessarily, finite, we must replace this with a condition at $x=1$. We choose to use the condition $d\theta/dx= 0$ at this boundary, equivalently $y(1) = 0$. Although not strictly accurate, the assumption of a zero gradient is expected to lead to small error as long as the glass has cooled sufficiently and the viscosity has become sufficiently large that there is negligible deformation beyond $x = 1$.
%It is therefore important that we always check that this is indeed the case. Hence, the boundary conditions are
% \begin{equation}
%     \chi|_{x = 0} = 1, \quad \quad \theta|_{x = 0} = 1, \quad \quad y|_{x = 1} = 0,\quad \quad \alpha|_{x = 0} = \alpha_\text{in}.
%         \label{Eq:second_order_BC}
% \end{equation}
In addition we also must choose the tension parameter $\sigma$ in \eqref{Eq:second_chi} so as to satisfy the draw ratio at $x = 1$. Therefore, the boundary conditions at $x = 1$ are given by
\begin{equation}
    y|_{x = 1} = 0,\quad \quad \chi|_{x=1}=1/\sqrt{D}.
    \label{Eq:original_BC_one}
\end{equation}

% \begin{subequations}
%     \begin{align}
%         &\chi|_{x = 0} = 1, \\
%         & \theta|_{x = 0} = 1, \quad \theta|_{x \rightarrow \infty} = \theta_a,\\
%         & \alpha|_{x = 0} = \alpha_\text{in}.
%     \end{align}
%     \label{Eq:original_BC}%
% \end{subequations}
% \\

%Also, we have added the condition $d\theta/dx = 0$ once temperature $\theta$ reaches the ambient temperature to prevent further, non-physical, decrease of temperature. \\

\section{Numerical solution method}

While the above system of equations may appear straightforward to solve, this is far from the case. The retention of heat conduction in the problem results in a second-order derivative of temperature in the equations which, in turn, makes the system challenging to solve numerically. The essential reason for this is that the temperature equation \eqref{eq:temperature} admits solutions that grow exponentially with $x$.
% , as can be readily seen from consideration of the general solution of the simplified heat equation
% \begin{align*}
%     \frac{A}{\Pe}\frac{d^2\theta}{dx^2}-\frac{d \theta}{dx} = H,
% \end{align*}
% where $A=\chi^2$ and $H$ are assumed to be constant. 
A rapid growth in temperature has a significant nonlinear effect on the viscosity. This is because the viscosity has the form \eqref{Eq:viscosity_new} and so it varies dramatically with temperature and this means that the solution is very sensitive. Of course, a rapid increase in temperature is not physically meaningful for our problem, in which the glass cools, so that rapid growth of temperature must be eliminated/controlled by the boundary condition on $\theta$ at $x=1$.

One might think that this problem can be attacked by using a shooting method, formulating the problem as an initial value problem and searching for an initial value $y(0)$ (equivalently $d\theta/dx$ at $x=0$) that yields the required boundary condition at $x=1$, i.e. $y(1)=0$. However, for typical choices of $y(0)$ the temperature grows exponentially and the nonlinear coupling via the viscosity causes serious numerical problems. Thus, a shooting method is not appropriate.

One might also think the problem would be readily solved using a boundary value problem solver (for example, \textsc{Matlab} solvers bvp4c, bvp5c), However, these routines require initial guesses for the unknown variables and for the reasons explained above the solution is extremely sensitive to these. Hence, this too is deemed to be an unsatisfactory solution method.

In place of these methods we have devised a novel iterative finite difference numerical solution scheme that suppresses the unphysical growth in temperature.
We discretise our spatial domain $0\leq x\leq 1$ with $N$ uniformly spaced gridpoints, $x_j$, $j=1,2,\ldots, N$, such that $\Delta x=x_j-x_{j-1}$ is the separation between any two consecutive gridpoints. %We discretised the dependent variables $\chi$, $\theta$, $y$, and $\alpha$ over a uniform grid of the independent variable $x$, with $N$ grid points over $0 \leq x \leq 1$.
%For $N$ grid points over $0 \leq x \leq 1$, the interval $\Delta x$ between grid points is 
%\begin{equation}
%    \Delta x = \frac{1}{N-1}.
%\end{equation}
%Let $x_j$ be the $j$th grid point, where $j = 1$ corresponds to $x = 0$ and $j = N$ corresponds to $x = 1$. 
We define $\chi_j=\chi(x_j)$ as the value of $\chi$ at gridpoint $x_j$, and similarly for $\theta_j$, $y_j$ and $\alpha_j$. % to be the values of $\chi$, $\theta$, $y$ and $\alpha$ at the $j$th grid point. 
Then, noting that $\chi_1$, $\theta_1$ and $\alpha_1$ are known from the boundary conditions at $x=0$, we discretise the differential equations for $\chi$, $\theta$ and $\alpha$ using %For the equations with boundary condition at $x = 0$, which are those for $\chi$, $\theta$, and $\alpha$, 
the well-known forward-Euler method, enabling the next iterate of each of these unknown functions to be computed at each grid point %. was utilised to solve the discretised equations
step-by-step from $x = 0$ to $x = 1$ (left to right), starting with the known value for $j = 1$ and ending with $j=N$. Previous iterates of the dependent variables are used as required. %, which we denote $\chi_j^{(p)}$, $\theta_j^{(p)}$, $\alpha_j^{(p)}$.

Equation \eqref{Eq:second_y} for $y$, however, has a boundary condition at $x = 1$, i.e. the value $y_N$ is known. Therefore, we discretise \eqref{Eq:second_y} using the backward-Euler method. %, discretising the $y$ equation using backward Euler method. 
Usually 
%equation \eqref{eq:backward_euler_y_2} 
the discretised equation would be seen as an implicit equation for determining $y_{j+1}$ from $y_j$. %\citep{numericalanalysis}. 
However, we rearrange it to give an explicit equation for $y_j$, assuming known values or previous iterates for the dependent variables at $x_{j+1}$. Thus we compute in the reverse direction from right ($x=1$) to left ($x=0$), starting with the known value $y_N$ and ending with $y_1$. This is a non-conventional technique but %was found to 
works well %for different initial guesses of the solution 
for our problem, and enables the boundary condition at $x = 1$ to be easily included. %Thus, we start from $y_N = 0$ ($x=1$) as our $y_{j+1}$ for $j = N-1$ to solve for $y_{N-1}$. Then, we use $y_{N-1}$ to solve for $y_{N-2}$, and so on until we reach $y_1$ ($x = 0$). 

\begin{figure}
\centering
\begin{tikzpicture}[scale=0.75]
%\hspace*{3em}
  % labels
    % \draw (0,0.5) node [anchor=west, text width=3.5cm]{$y_1$};
    % \draw (14,0.5) node [anchor=west, text width=3.5cm]{$y_N = 0$};
      
  \draw (0,-0.5) node [anchor=west, text width=3.5cm]{$x_1 = 0$};
  \draw (14,-0.5) node [anchor=west, text width=3.5cm]{$x_N = 1$};
  
    \draw[<->] (6.75,0.5) -- (8.75,0.5) node [midway,fill=white]{$\Delta x$};
      \draw[ultra thick,<-] (2.75,1) -- (12.75,1);
      \draw(7.75,1.1) node [anchor=south]{{\small\fbox{2}} $y_1\leftarrow \ldots\leftarrow y_{N-1}\leftarrow y_N$};
      \draw[ultra thick,->] (2.75,-1) -- (12.75,-1);
      \draw(7.75,-1.1) node [anchor=north]{{\small\fbox{1}} $\chi_1\rightarrow\chi_2\rightarrow\ldots\rightarrow\chi_N$, {\small\fbox{3} }$\theta_1\rightarrow\theta_2\rightarrow\ldots\rightarrow\theta_N$, {\small\fbox{4}}
      $\alpha_1\rightarrow\alpha_2\rightarrow\ldots\rightarrow\alpha_N$};
      %\draw(7.75,-1.5) node [anchor=north]{$\theta_1,\theta_2,\ldots,\theta_N$}

\draw (0.75,0) -- (14.75,0);
\draw (0.75,-0.25) -- (0.75,0.25);
\draw (2.75,-0.25) -- (2.75,0.25);
\draw (4.75,-0.25) -- (4.75,0.25);
\draw (6.75,-0.25) -- (6.75,0.25);
\draw (8.75,-0.25) -- (8.75,0.25);
\draw (10.75,-0.25) -- (10.75,0.25);
\draw (12.75,-0.25) -- (12.75,0.25);
\draw (14.75,-0.25) -- (14.75,0.25);

\end{tikzpicture}
\caption{Finite difference solution method; $\chi_j$, $\theta_j$, $\alpha_j$, are computed from left to right, i.e. for $j=2,3,\ldots,N$ given the known values $\chi_1,\ \theta_1,\ \alpha_1$,  while $y_j$ is computed from right to left, i.e. for $j=N-1,N-2,\ldots,1$ given $y_N$. The boxed numbers show the order of computation of the unknown variables.}
\label{fig:diagram_backward}
\end{figure}

%The backward Euler formula was used to discretise the differential equation for $y$.
%\eqref{Eq:second_y}.
% ; that is
% \begin{equation}
%     \frac{dy_{j+1}}{dx} =\frac{y_{j+1}-y_j}{\Delta x}.
%     \label{eq:backward_euler_y}
% \end{equation}
% This was substituted for the left hand side of equation \eqref{Eq:second_y}. Discretising the right hand side and remembering we are evaluating at $x_{j+1}$, we obtain
% \begin{equation}
%     \frac{y_{j+1}-y_j}{\Delta x} = \left(\frac{y_{j+1}\Pe }{\chi_{j+1}^2} + \chi_{j+1}\left(\frac{1}{{\alpha}_{j+1}}+\pi {\alpha}_{j+1}\right) \left[\HR\left({\theta}_{j+1}^4 - \theta_a^4\right) 
%      + \HC \left({\theta}_{j+1}^4 - \theta_a^4\right)\right]\right).
%     \label{eq:backward_euler_y_2}
% \end{equation}

%Using the forward and the backward Euler formulas as explained above gives the following system of discrete equations and boundary conditions, searching for a tension to achieve the draw ratio at $x = 1$:
We thus obtain the following system of difference equations to be iteratively evaluated for a given tension $\sigma$ until the change in the dependent variables becomes less than some required tolerance:
\begin{subequations}
   \begin{align}
        \chi_{j+1} &= \chi_j + \Delta x \left(\frac{\chi_j^2}{6 {\alpha}^{(p)}_{j}\Ca \mu\left(\theta^{(p)}_{j}\right)} - \frac{\chi_j \sigma}{\mu\left(\theta^{(p)}_{j}\right)}\right) \text{, } \label{Eq:finite_chi}\\
          y_{j} &= y_{j+1}-\Delta x \left(\frac{y_{j+1}\Pe }{\chi^{2}_{j+1}} + \chi_{j+1}\left(\frac{1}{\alpha^{(p)}_{j+1}}+\pi \alpha^{(p)}_{j+1}\right) \left[\HR\left({\theta^{(p)}_{j+1}}^{4} - \theta_a^4\right) \right. \right. \nonumber \\
          &\left. \left. + \HC \left(\theta^{(p)}_{j+1} - \theta_{a}\right) \right] \right) \text{,} \label{finite_y}\\ 
         \theta_{j+1} &= 
         \begin{dcases}
            \theta_j + \Delta x \frac{\Pe}{\chi^{2}_{j}}y_{j} \text{ , } &\theta_j > \theta_a,\\
            \theta_j \text{ , } &\theta_j \leq \theta_a,
        \end{dcases}\label{Eq:finite_theta}  \\ 
         \alpha_{j+1} &=
        \begin{dcases}
        \alpha_j + \Delta x \left(\frac{\chi_j}{2\Ca \mu\left(\theta_j\right)}\right) \text{, } &\alpha < 1/\sqrt{\pi},\\
        \alpha_j \text{, } &\alpha \geq 1/\sqrt{\pi},
    \end{dcases}\label{Eq:finite_alpha}  
        \end{align}
        \label{Eq:discretised}
    \end{subequations}
where a supercript $(p)$ denotes the previous iterate of a variable.

%Each equation in \eqref{Eq:second_order} can be solved independently assuming all other unknowns are known. Starting with an initial guess of the solution, this can then be updated iteratively until convergence to the solution is achieved. \\
%Note that $\chi_{j}^{(p)}, \theta_{j}^{(p)},$ and $\alpha_{j}^{(p)}$ denote values of these variables at the previous iteration. 
Our numerical finite-difference method is illustrated in Figure~\ref{fig:diagram_backward}. A Gauss-Seidel iterative scheme is employed, making use of updated values of variables as soon as they are known, so that only previous iterates of $\theta_j$ and $\alpha_j$ are required. To start the iterative process, initial guesses for these are needed and these were chosen %to be functions to approximate the solutions given in the first-order problem, 
as $\theta_j^{(p)}=(1-x_j)^2$ and $\alpha_j^{(p)}=\alpha_\text{in}$. In addition we require %With these and 
the boundary conditions,
    \begin{equation}
        \chi_1 = 1, \quad \theta_1 = 1, \quad y_N = 0, \quad \alpha_1 = \alpha_\text{in}.
            \label{Eq:discretised_BC}%
    \end{equation}
   %the first iterate of the solution is found. 
   For robustness, we solve  the equations in the order given in \eqref{Eq:discretised}. %of $\chi$, $y$, $\theta$, then $\alpha$ as indicated in equation . 
In particular, it is important to include the boundary condition at $x = 1$ by solving for $y$ before solving for $\theta$. Solving for $\theta$ before $y$ can result in numerical difficulties when solving for $\theta$. %Then, the Gauss--Seidel iteration scheme which uses updated values of variables as soon as they become available, was used. 
To test for convergence, we compute the magnitude of the %use the absolute 
difference between two consecutive iteratives for each dependent variable and require that the maximum of these be less than % and compare it with 
a chosen tolerance value. The tolerance value was set at $10^{-8}$, giving a good balance between speed and accuracy of the solution.

Finally, the above procedure was written as a function to be used with \textsc{Matlab}'s root-finding routine `fzero' so as to find the tension $\sigma$ corresponding to a desired draw ratio $D=1/\chi_N^2$. 

Having obtained a solution $\chi_j(x)$, $\theta_j(x)$, $\alpha_j(x)$ we may compute the aspect ratio $\phi_j(x)$ and external radius $R_j(x)$ of the tube from \eqref{Eq:alpha_phi} and \eqref{eq:chi_R}, respectively.

\section{Results}
% For our purposes in this paper, we assume unit heat transfer coefficients and capillary number, i.e. $\HR=1$, $\HC=1$, $\Ca=1$. %This paper assumes that the heat transfer coefficients are $\HR = 1$ and $\HC = 1$, and also that the capillary number is $\Ca = 1$ for exploratory purposes. 
% The effect of each of the heat transfer coefficients is explored in \cite{eunice_thesis}, where it is shown that they %Both of the heat transfer coefficients 
% have a similar effect on the solution of the problem. 
In this paper we focus on the importance of conductive heat transport in the pulled extrusion process, as quantified by the P\'eclet number $\Pe$. %We analyse the effect of conductive heat transport, represented by the $\Pe$ term in the temperature equation. 
In the limit $\Pe \rightarrow \infty$, the second-order term in the temperature equation may be neglected, reducing it to first-order. 
%The first-order problem is obtained by assuming a very large P\'eclet number, $\Pe \rightarrow \infty$, leading to neglect of the second-order derivative of the temperature $\theta$. 
We will explore how different values of  $\Pe$ %P\'eclet number 
affect the %second-order problem 
solution. When the second-order term in the temperature equation is neglected we describe the coupled energy and flow extrusion problem as the ``first-order problem''; otherwise it is described as the ``second-order problem''. Below we will, at times, use subscripts $i=1,2$ on quantities $X$, where $X_1$ denotes the quantity as given by solving the first-order problem and $X_2$ denotes the quantity as given by solving the second-order problem.  % and the subscript 2 to denote quantities re  or 2 denotes the first- or second-order problems, respectively. \\

Aside from the P\'{e}clet number, all other parameters are set to fixed values. We choose the constant $C = 10$ in the dimensionless viscosity-temperature function \eqref{Eq:viscosity_new}. This parameter governs the sensitivity of the viscosity to temperature and depends on the glass and the temperature range in the extrusion process. In practice, this value could be significantly larger. We further choose a representative draw ratio $D=2$, initial aspect ratio $\phi_\text{in}=0.2$ (equivalently $\alpha_\text{in}=0.4607$), and dimensionless ambient air temperature $\theta_a = 0.2990$. We also assume unit heat transfer coefficients and capillary number, i.e. $\HR=1$, $\HC=1$, $\Ca=1$. For investigations of these last three parameters see \cite{eunice_thesis}.

 % We search for the tension that gives a draw ratio of $D = 2$. We denote this tension $\sigma_2$ and compare this with the value $\sigma_1$ obtained when using the first-order temperature equation. The dimensionless ambient air temperature is $\theta_a = 0.2990$ from Table \ref{tab:parameters_val}. The ratio of inner to outer radius initially is chosen to be $\phi_\text{in} = 0.2$, giving the scaled wall thickness $\alpha_\text{in} = 0.4607$. Also, we choose the constant $C = 10$ in the viscosity-temperature function for the analysis of the model in this paper. The dimensionless parameter $C$ governs the sensitivity of the viscosity to temperature and depends on the glass and the temperature range in the extrusion process. We have not attempted to match to a particular extrusion but simply chosen $C = 10$. In practice this value could be significantly larger. 

\begin{figure}
     \centering
     \begin{subfigure}[b]{0.45\textwidth}
         \centering
         \includegraphics[width=\textwidth]%{the_thesis/Figures/theta_Pe_10.jpg}
         {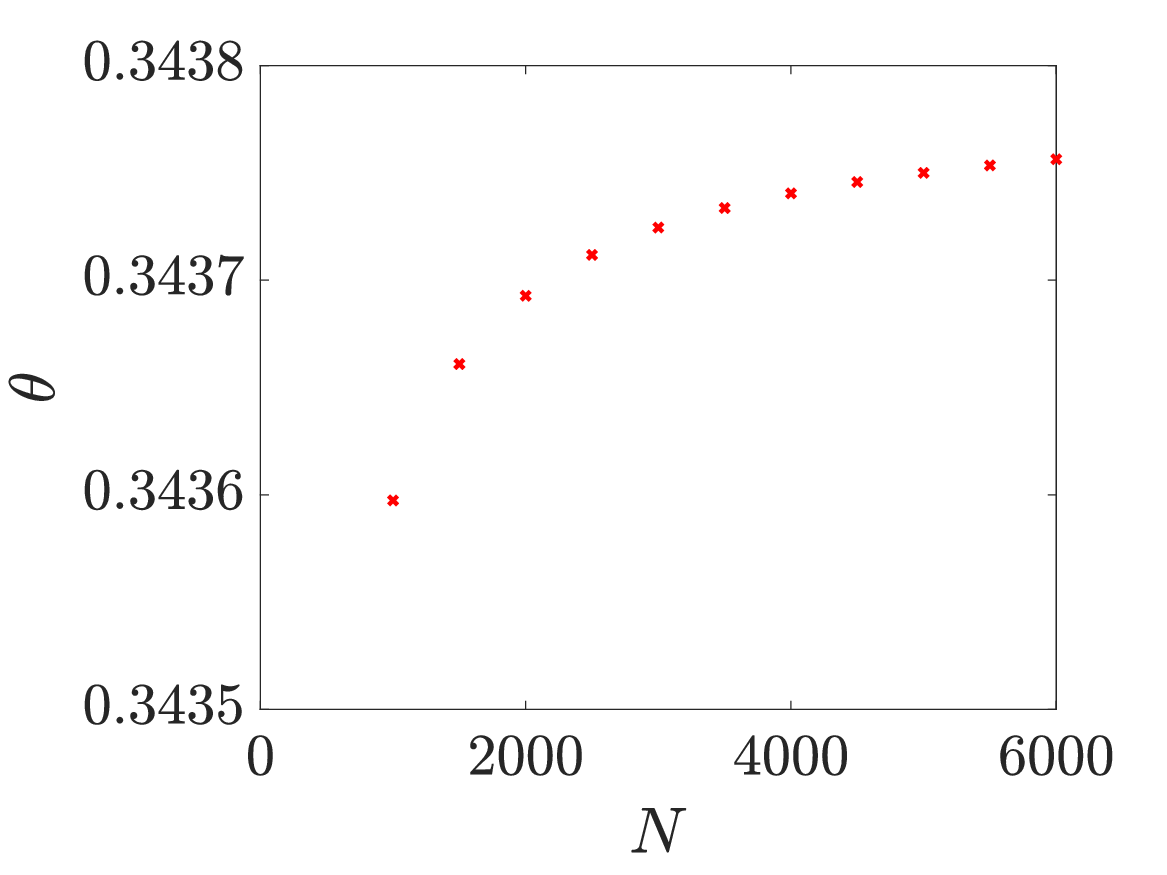}
         \caption{$\theta_2(1)$, $\Pe = 10$}
         \label{fig:conv_theta_10}
     \end{subfigure}
          \begin{subfigure}[b]{0.45\textwidth}
         \centering
         \includegraphics[width=\textwidth]%{the_thesis/Figures/theta_Pe_600.jpg}
         {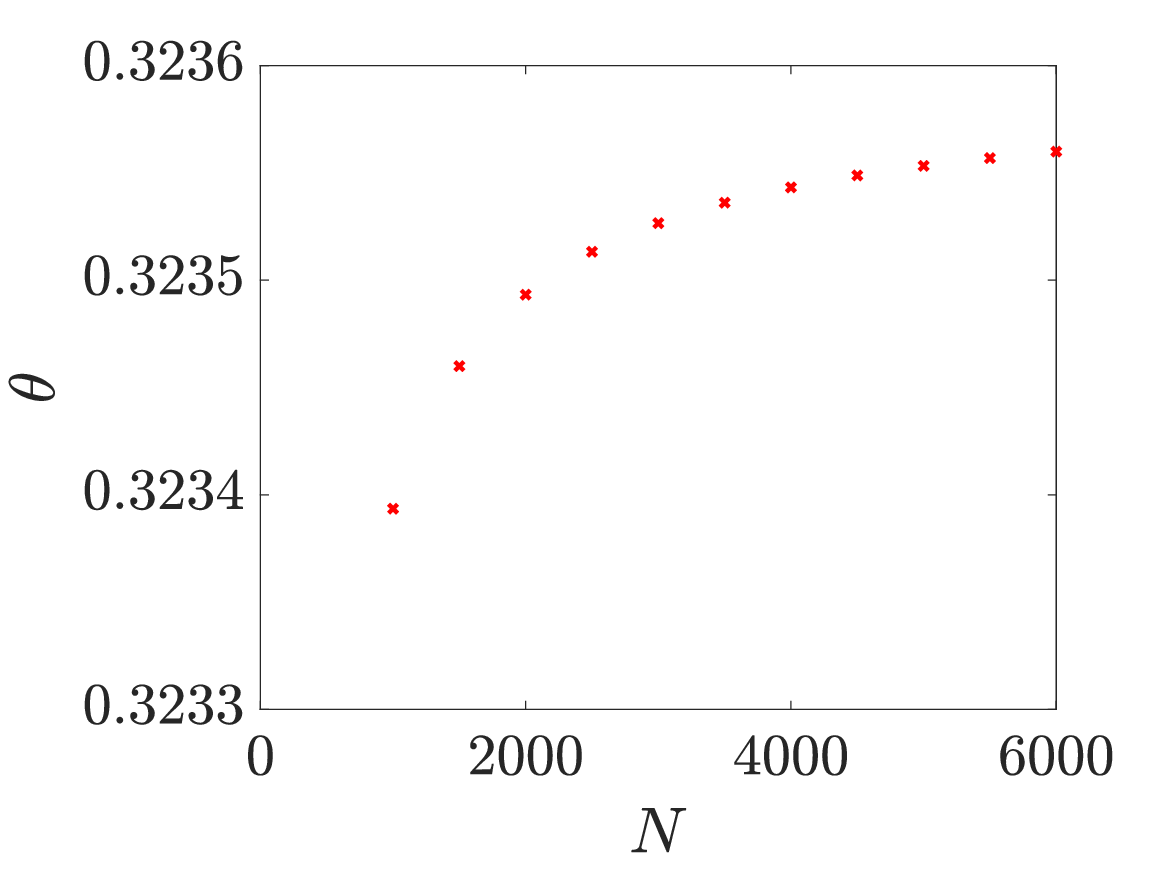}
         \caption{$\theta_2(1)$, $\Pe = 600$}
         \label{fig:conv_theta_500}
     \end{subfigure}
          \begin{subfigure}[b]{0.45\textwidth}
         \centering
         \includegraphics[width=\textwidth]%{the_thesis/Figures/phi_Pe_10.jpg}
         {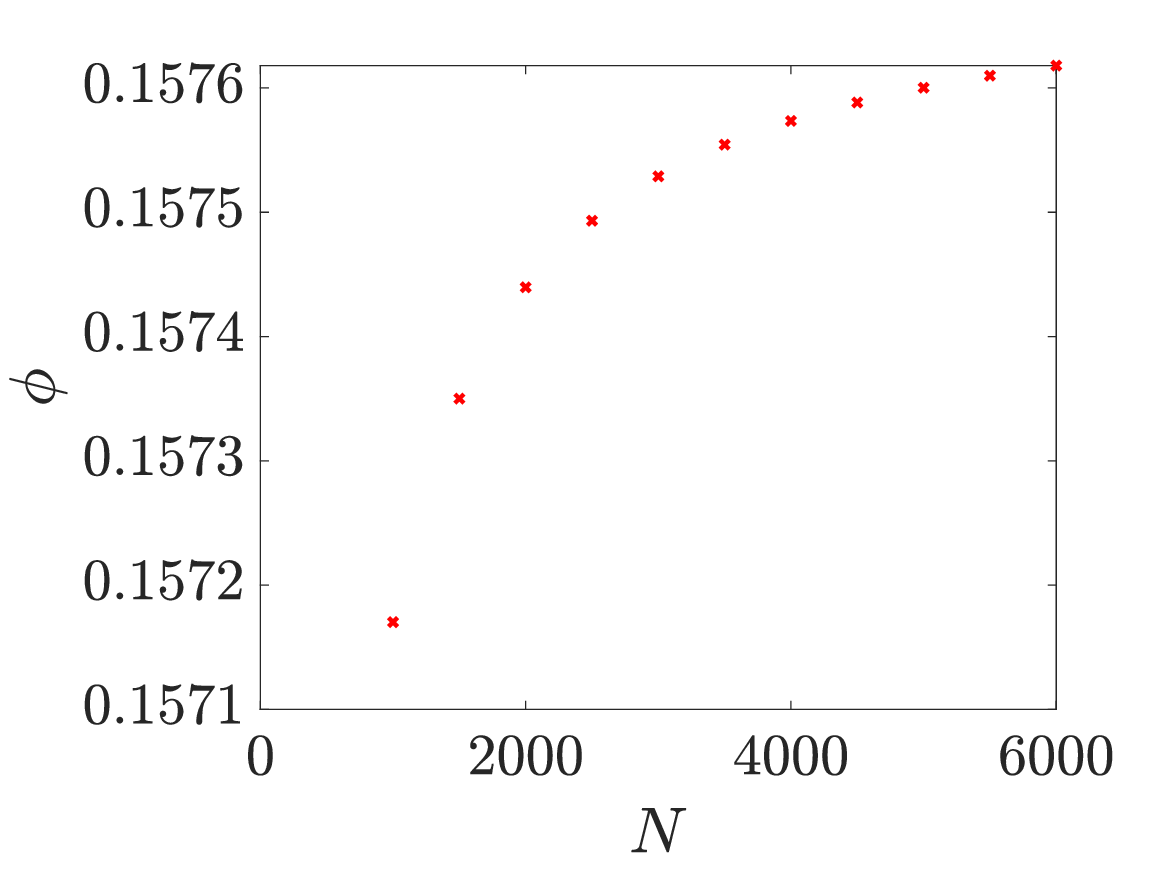}
         \caption{$\phi_2(1)$, $\Pe = 10$}
         \label{fig:conv_phi_10}
     \end{subfigure}
     \begin{subfigure}[b]{0.45\textwidth}
         \centering
         \includegraphics[width=\textwidth]%{the_thesis/Figures/phi_Pe_600.jpg}
         {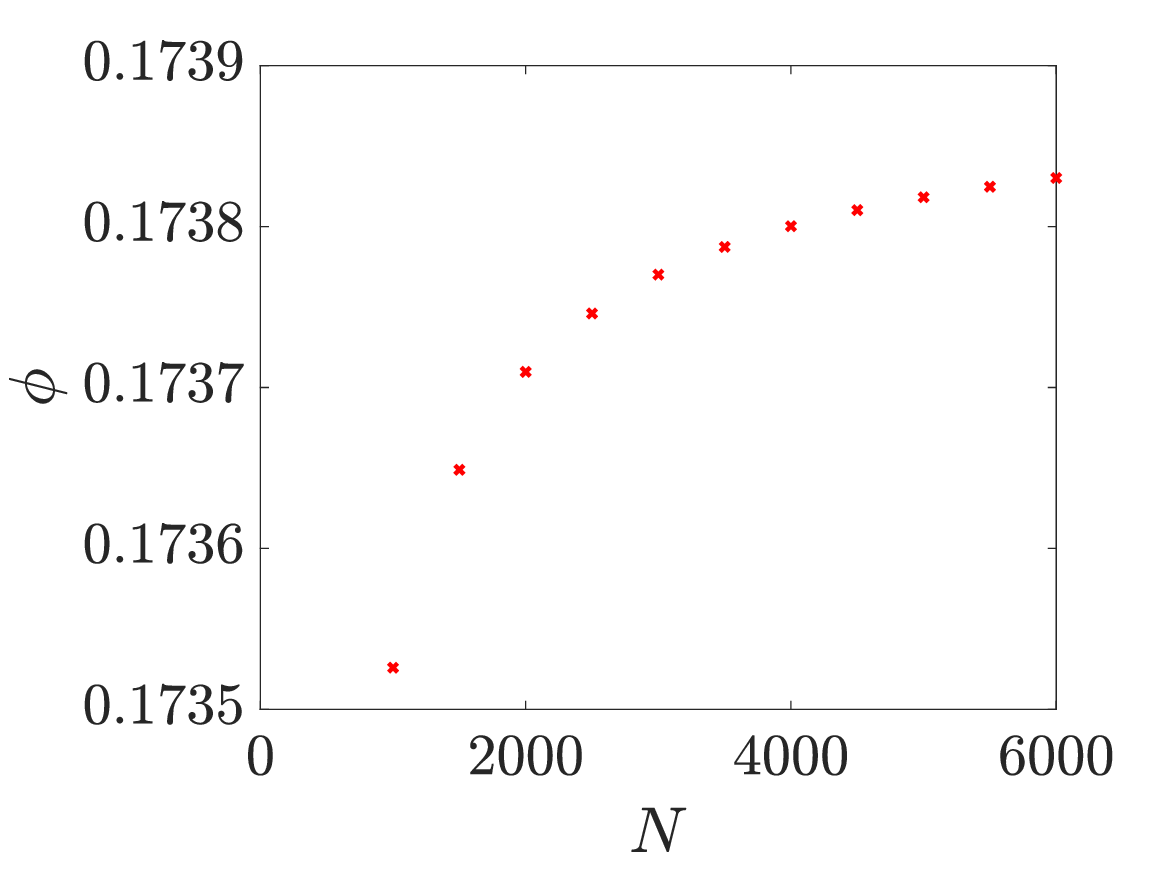}
         \caption{$\phi_2(1)$, $\Pe = 600$}
         \label{fig:conv_phi_500}
     \end{subfigure}
     \begin{subfigure}[b]{0.45\textwidth}
         \centering
         \includegraphics[width=\textwidth]%{the_thesis/Figures/sigma_Pe_10.jpg}
         {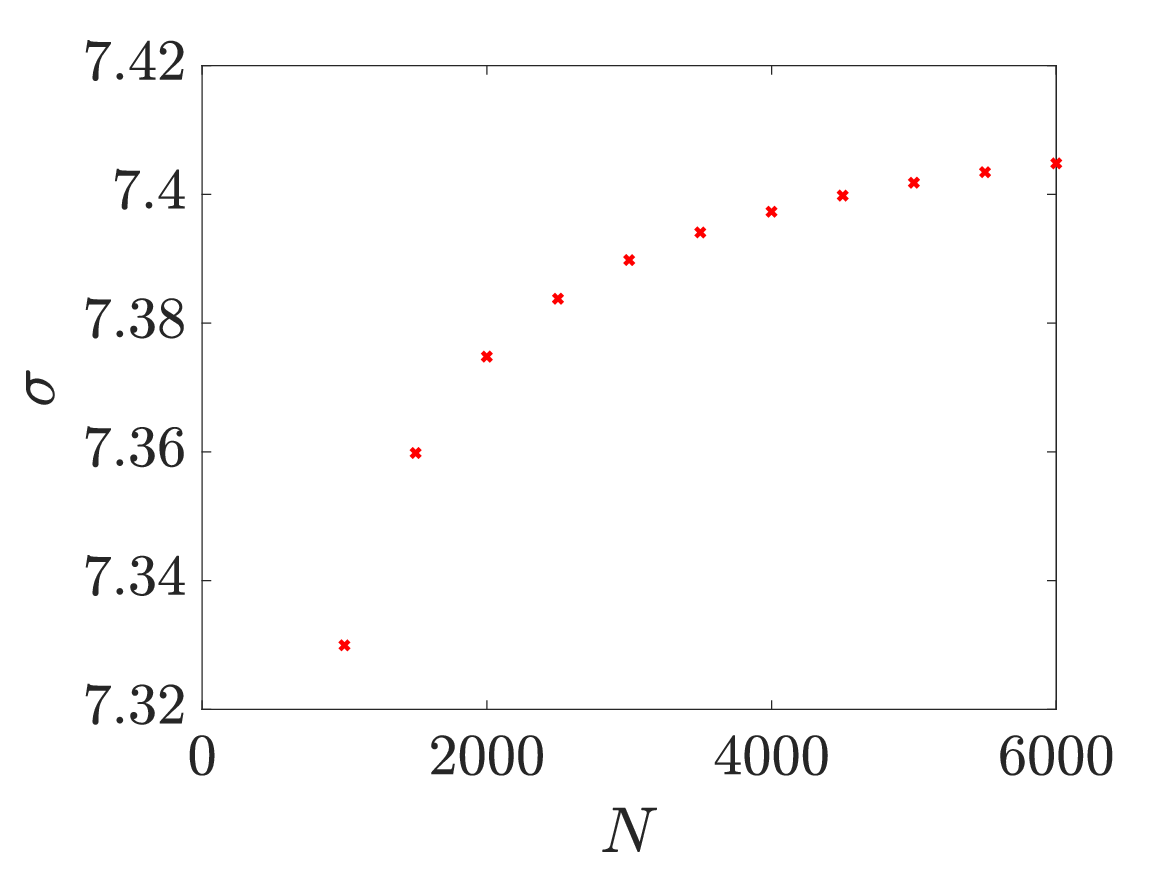}
         \caption{$\sigma_2$, $\Pe = 10$}
         \label{fig:conv_sigma_10}
     \end{subfigure}
     \begin{subfigure}[b]{0.45\textwidth}
         \centering
         \includegraphics[width=\textwidth]%{the_thesis/Figures/sigma_Pe_600.jpg}
         {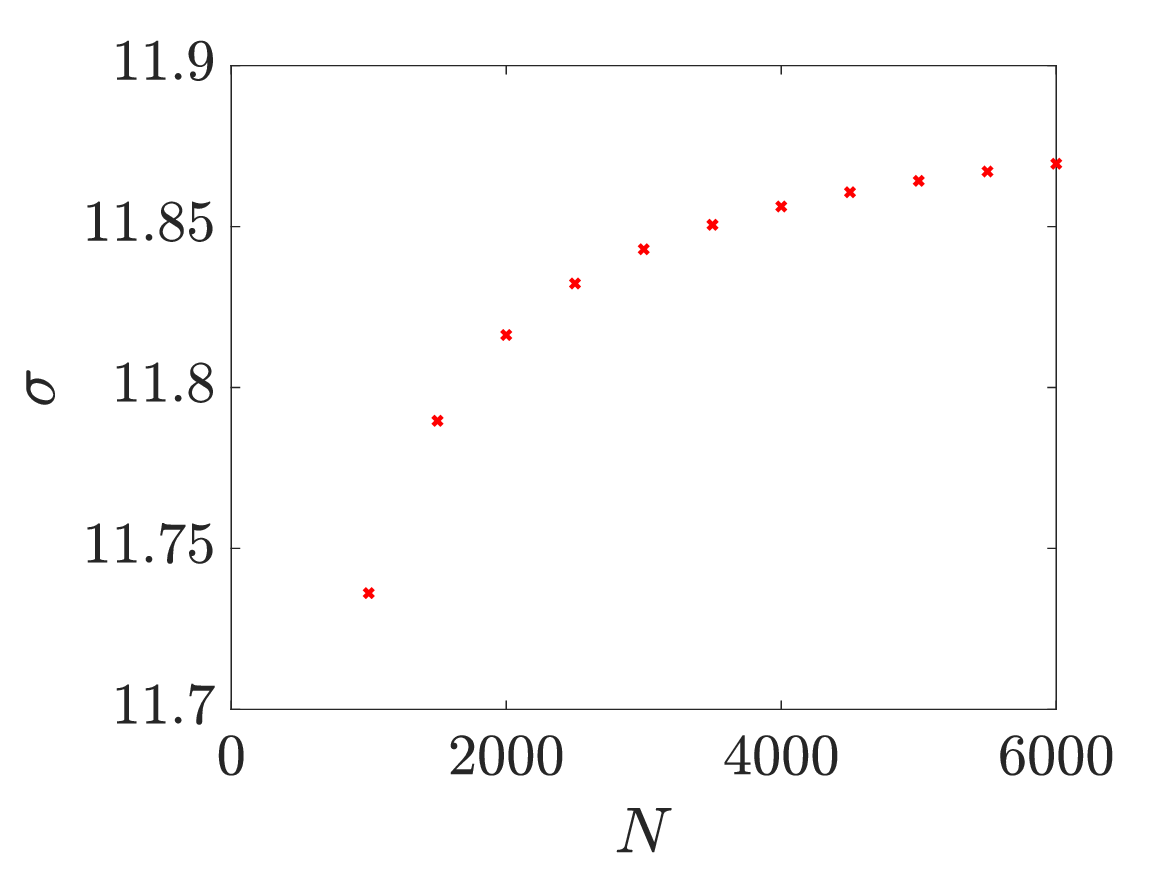}
         \caption{$\sigma_2$, $\Pe = 600$}
         \label{fig:conv_sigma_500}
     \end{subfigure}
        \caption[Convergence plots, varying number of grid points $N$.]{Convergence plots of $\theta_2(1)$ (first row), $\phi_2(1)$ (second row), and $\sigma_2$ (third row) against the number of grid points $N$ for $\Pe = 10$ (first column) and $\Pe = 600$ (second column). Other parameters: $D = 2$, $\Ca = 1$, $\HR = 1$, $\HC = 1$, $C=10$, $\phi_\text{in}=0.2$, $\theta_a=0.2990$.}
        
 \label{fig:convergence}
\end{figure}

%We need to choose the number of grid points $N$ for our finite-difference discretisation. %to utilize the finite difference method. 
To determine the number of grid points $N$ for our finite-difference discretisation, such that our solution is sufficiently accurate, we computed solutions % suitable for this problem, the second-order solution for temperature $\theta_2(1)$, aspect ratio $\phi_2(1)$, and tension $\sigma_2$ are plotted against the number of grid points $N$ 
for two different values of the P\'eclet number, $\Pe=10,\,600$, for different values of $N$. % of 10 and 600 in 
Figure \ref{fig:convergence} shows the temperature $\theta_2(1)$, aspect ratio $\phi_2(1)$, and pulling tension $\sigma_2$ versus $N$. As seen there, the second-order solution converges as $N$ increases; increasing from $N=5000$ to $N=6000$ gives changes to these quantities in %, with variations of 
the fourth and fifth decimal places. % on increasing from $N = 5000$ to $N = 6000$.
Therefore, we choose $N = 5000$, which provides a good balance between accuracy and computational cost. 

\begin{figure}
     \centering
     % \begin{subfigure}[b]{0.4\textwidth}
     %     \centering
     %     \includegraphics[width=\textwidth]{the_thesis/Figures/chi_five.jpg}
     %     \caption{$\chi(x)$}
     %     \label{fig:chi_Pe}
     % \end{subfigure}
          \begin{subfigure}[b]{0.495\textwidth}
         \centering
        \includegraphics[width=\textwidth]%{Figures/trial_temp.eps}
        {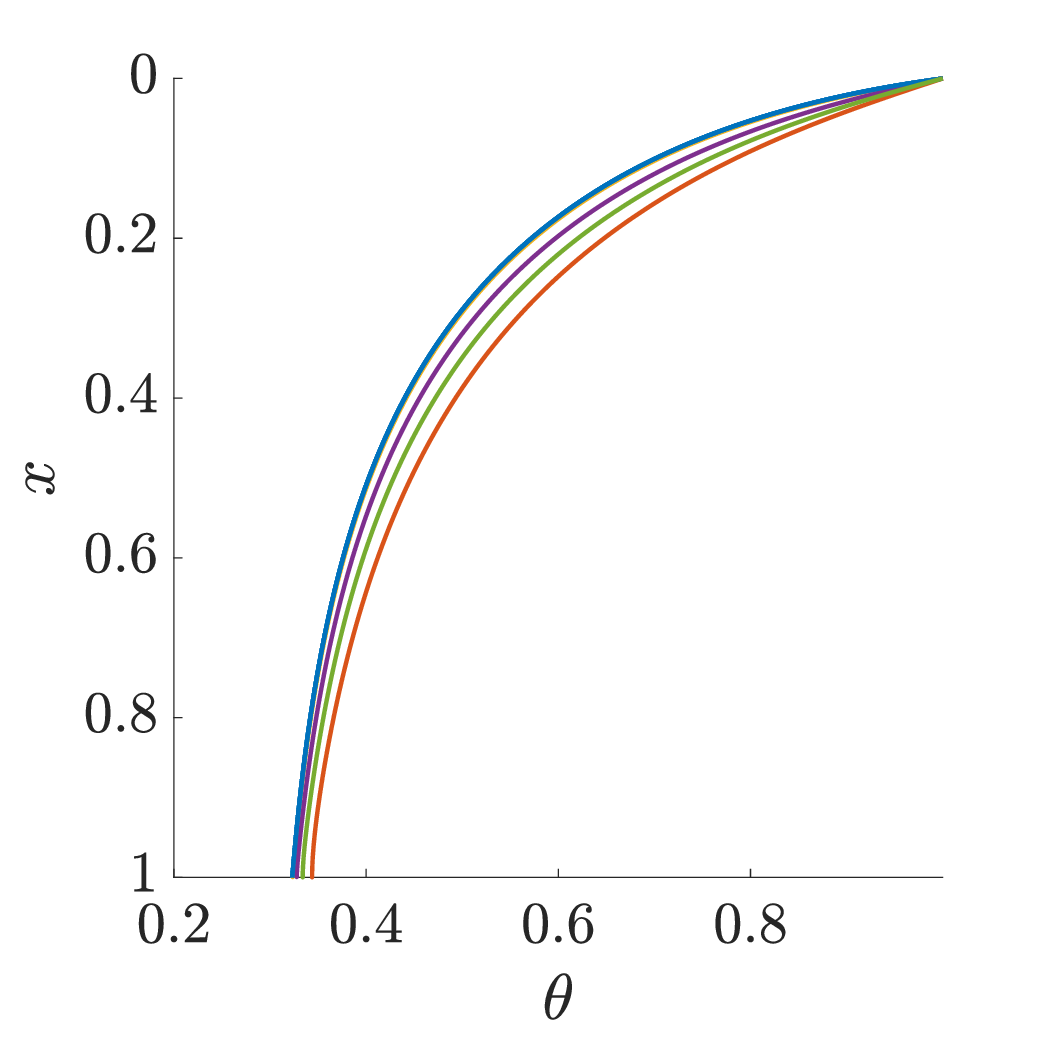}
         \caption{$\theta(x)$}
         \label{fig:theta_Pe}
     \end{subfigure}
     %      \begin{subfigure}[b]{0.4\textwidth}
     %     \centering
     %     \includegraphics[width=\textwidth]{the_thesis/Figures/alpha_five.jpg}
     %     \caption{$\alpha(x)$}
     %     \label{fig:alpha_Pe}
     % \end{subfigure}
     \begin{subfigure}[b]{0.495\textwidth}
         \centering
         \includegraphics[width=\textwidth]%{the_thesis/Figures/phi_five.jpg}
         {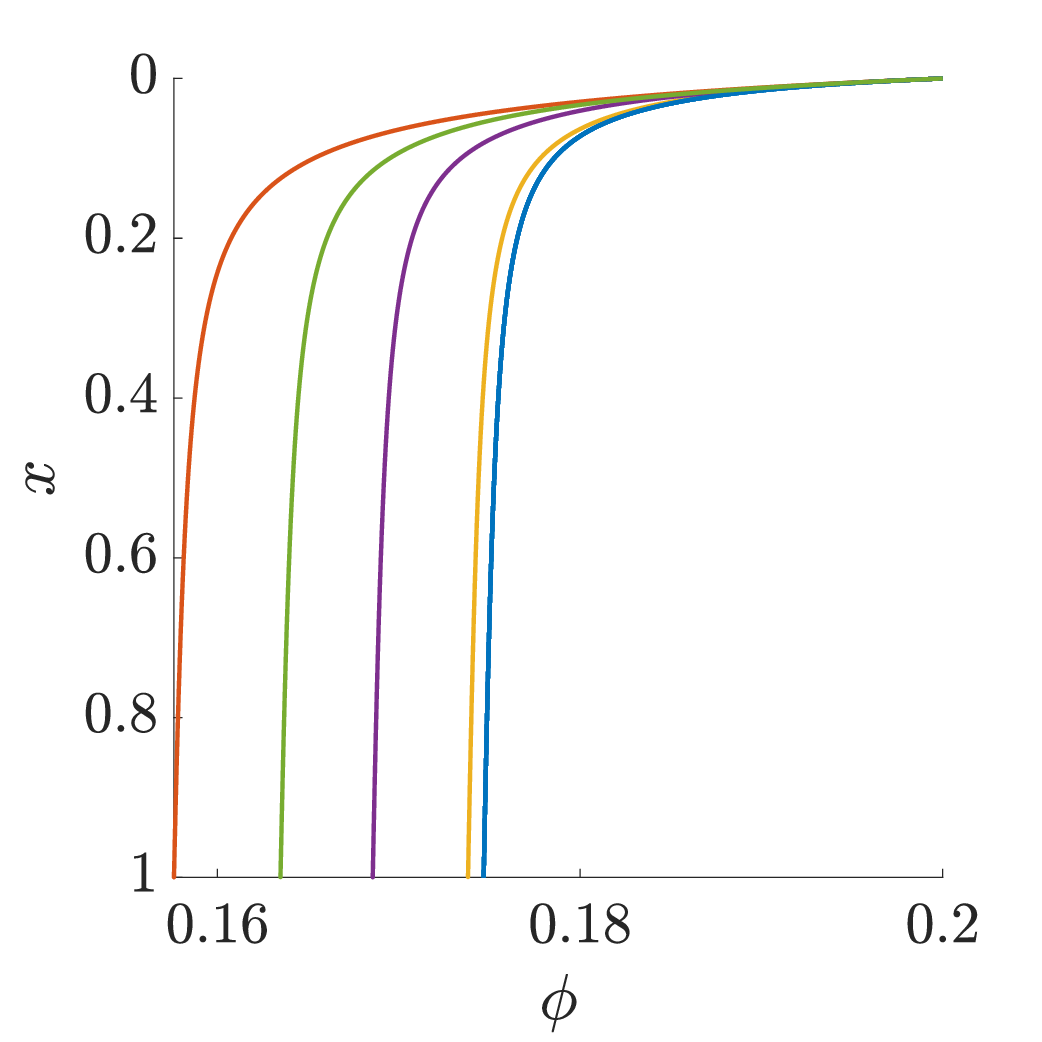}
         \caption{$\phi(x)$}
         \label{fig:phi_Pe}
     \end{subfigure}
     \begin{subfigure}[b]{0.495\textwidth}
         \centering
         \includegraphics[width=\textwidth]%{the_thesis/Figures/R_five.jpg}
         {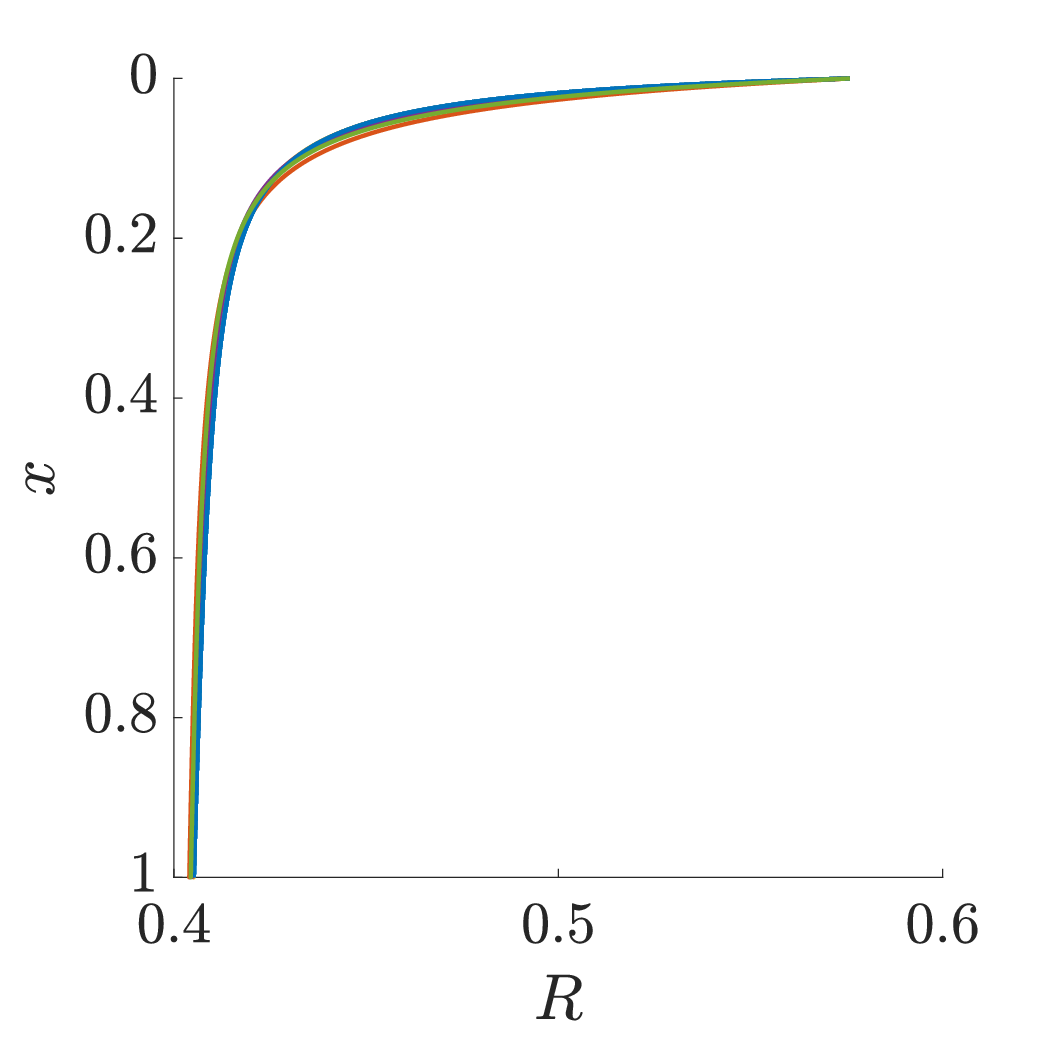}
         \caption{$R(x)$}
         \label{fig:R_Pe}
     \end{subfigure}
     % \begin{subfigure}[b]{0.4\textwidth}
     %     \centering
     %     \includegraphics[width=\textwidth]{the_thesis/Figures/phiR_five.jpg}
     %     \caption{$\phi(x) R(x)$}
     %     \label{fig:phiR_Pe}
     % \end{subfigure}
        \caption[First-order and second-order solution obtain using parameters $D = 2$, $\Ca = 1$, $\HR = 1$, and $\HC = 1$.]{(a) Temperature $\theta(x)$, (b) aspect ratio $\phi(x)$, (c) external radius $R(x)$. First-order solution $\Pe \rightarrow \infty$ (blue curve), and second-order solutions $\Pe = 10$ (red curve), $\Pe = 20$ (green curve), $\Pe = 50$ (purple curve), $\Pe = 600$ (yellow curve). Other parameters: $D = 2$, $\Ca = 1$, $\HR = 1$, $\HC = 1$, $C=10$, $\phi_\text{in}=0.2$, $\theta_a=0.2990$.}
        \label{fig:Pe}
\end{figure}

%Figure \ref{fig:Pe} shows $\chi(x)$, $\theta(x)$, $\alpha(x)$, $\phi(x)$, $R(x)$, and $\phi(x)R(x)$ plotted against $x$ for $\Pe \rightarrow \infty $ (blue), $\Pe = 10$ (red) and $\Pe = 600$ (yellow). Looking at Figure \ref{fig:chi_Pe}, the solution profile for $\chi(x)$ is similar for all three cases with $\chi$ decreasing as $x$ increases. Initially, all of the solutions start with $\chi(0) = 1$ as prescribed in the model. The final value $\chi(1) = 1/\sqrt{2}$ is also prescribed since we are searching for the tension that gives a draw ratio of $D = 2$. The yellow curve is very close to the blue curve, indicating that the solution with high P\'eclet number is close to the first-order solution. \\

Figure \ref{fig:Pe} shows the temperature $\theta(x)$, aspect ratio $\phi(x)$, and external radius $R(x)$ of the tube plotted against $x$ for $\Pe \rightarrow \infty $ (blue), $\Pe = 10$ (red), $\Pe = 20$ (green), $\Pe = 50$ (purple), and $\Pe = 600$ (yellow). %The aspect ratio $\phi(x)$ is calculated using the $\alpha(x)$ solution from the numerical method and the relationship between $\alpha(x)$ and $\phi(x)$ of \eqref{Eq:alpha_phi}. The radius $R(x)$ is calculated from the $\chi(x)$ solution and the relationship \eqref{eq:chi_R}.
The temperature of the first-order problem (blue curve) is quite different from the second-order problem with lowest P\'eclet number, as shown by the red curve in Figure \ref{fig:theta_Pe}. However, as the P\'eclet number increases to 600 (yellow curve), the second-order solution approaches the first-order solution. % as shown by the yellow curve in the same figure.
%We also saw this in Table \ref{tab:varying_Pe}. 
Figure \ref{fig:theta_Pe} shows that heat conduction in the second-order problem with low P\'eclet number results in a higher temperature at $x = 1$. Indeed, for $\Pe = 10$, the temperature throughout the glass is higher than for large P\'eclet number. This is due to the reciprocal of the P\'eclet number being the coefficient of the conduction term. When the P\'eclet number is small, the conduction of heat in the glass is important, increasing the glass temperature.

%For the scaled wall thickness $\alpha(x)$ and the aspect ratio $\phi(x)$, we again see in Figures \ref{fig:alpha_Pe} and \ref{fig:phi_Pe} that the yellow curve (second-order solution with $\Pe = 600$) and blue curve (first-order solution) have similar solution profiles. However, the red curves show $\alpha(x)$ to be larger and $\phi(x)$ to be smaller over the domain $0 \leq x \leq 1$. These curves correspond to the second-order solution with low P\'eclet number $\Pe = 10$, for which heat conduction is having a significant effect on the the geometry. This relates to the higher glass temperature in the $\Pe = 10$ case resulting in more deformation and hence a thicker wall and a smaller aspect ratio for at any $x$. \\
%This is also shown for $x = 1$ in Table \ref{tab:varying_Pe}. \\

For the aspect ratio $\phi(x)$, we again see in Figure \ref{fig:phi_Pe} that the yellow curve (second-order solution with $\Pe = 600$) and blue curve (first-order solution) are very similar. %have similar solution profiles. 
However, the red, green, and purple curves show $\phi(x)$ to be significantly smaller for the smaller values of $\Pe$ %the yellow curve 
over the domain $0 < x \leq 1$, with $\phi$ increasing as $\Pe$ increases. The red curve corresponds to the second-order solution with lowest P\'eclet number, $\Pe = 10$, and shows that heat conduction has a significant effect on the geometry. This relates to the higher glass temperature in this case, resulting in more deformation and hence a thicker wall and a smaller aspect ratio at any $x>0$. 
%This is also shown for $x = 1$ in Table \ref{tab:varying_Pe}. \\

%The external radius $R$ in Figure \ref{fig:R_Pe} and the internal radius $\phi R$ in Figure \ref{fig:phiR_Pe} both decrease as $x$ increases. This is due to the thinning of the glass when it is pulled to achieve a draw ratio of $D = 2$. Similar to $\chi$, there is not much difference between the three curves in the solution profile for $R$. However, the inner radius $\phi(x) R(x)$ shows greater difference, with a solution profile similar to $\phi(x)$. Again, as the P\'eclet number increases from 10 (red curve) to 600 (yellow curve), the second-order solution approaches the first-order solution (blue curve). \\ 

The external radius $R$ in Figure \ref{fig:R_Pe} decreases as $x$ increases. This is due to the thinning of the glass when it is pulled to achieve a draw ratio of $D = 2$. There is not, however, much difference between the three solution curves for $R$; this is because the pulling tension $\sigma$ has been varied to achieve the same draw ratio for all three cases. Again, as the P\'eclet number increases from 10 (red curve) to 600 (yellow curve), the second-order solution approaches the first-order solution (blue curve). 
 
\begin{figure}
\centering
    \begin{subfigure}[b]{0.5\textwidth}
    \centering
    \includegraphics[width=\textwidth]%{the_thesis/Figures/vary_Pe_10_theta.jpg}
    {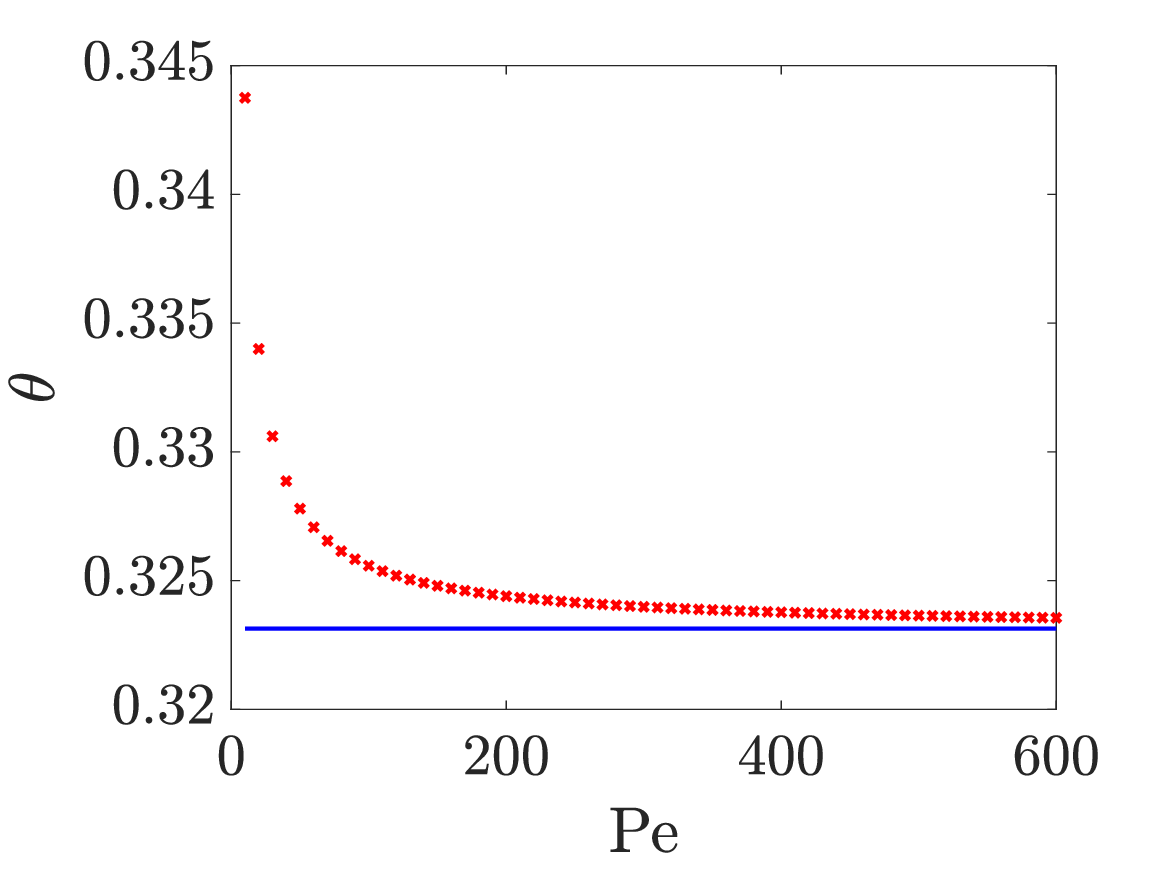}
    \caption{$\theta(1)$}
    \label{Fig:vary_Pe_theta}
    \end{subfigure}
    \begin{subfigure}[b]{0.5\textwidth}
    \includegraphics[width=\textwidth]%{the_thesis/Figures/vary_Pe_10_phi.jpg}
    {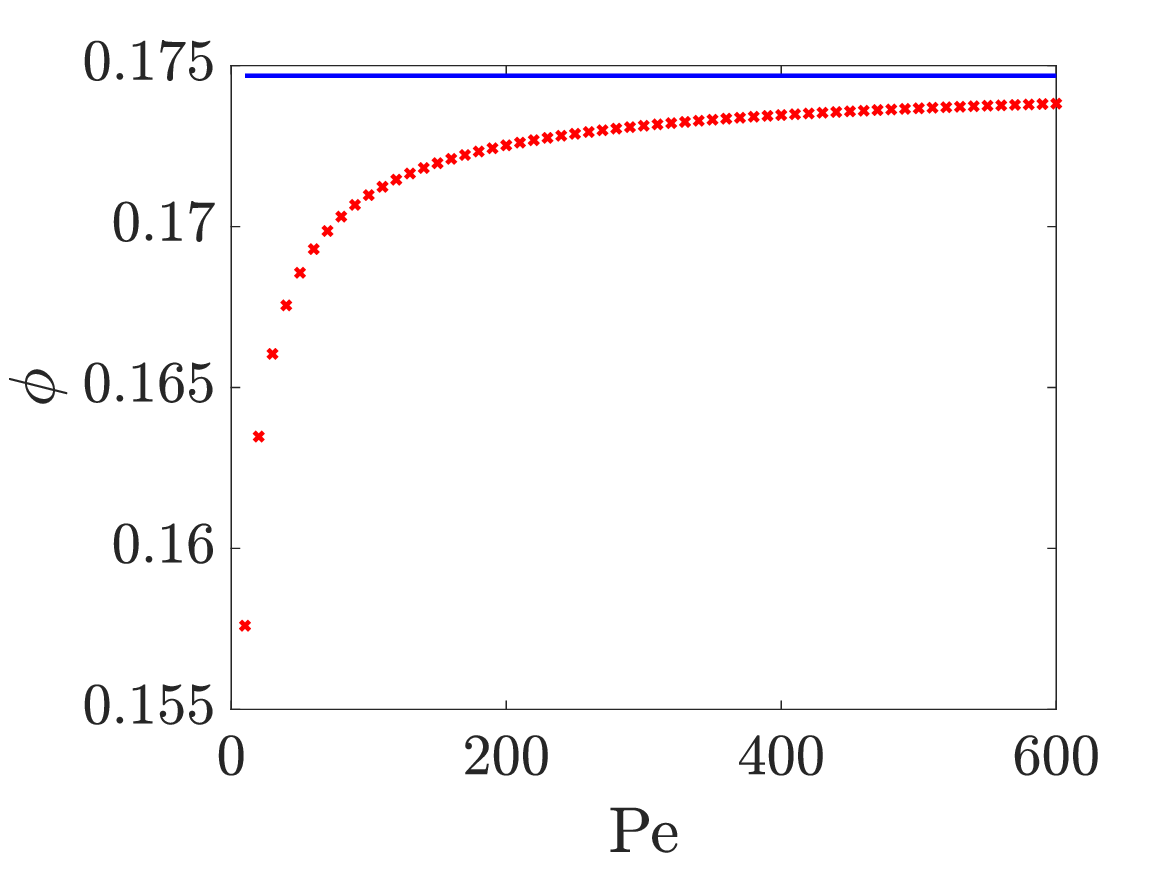}
    \caption{$\phi(1)$}
    \label{Fig:vary_Pe_phi}
     \end{subfigure}
    \begin{subfigure}[b]{0.5\textwidth}
    \centering
    \includegraphics[width=\textwidth]%{the_thesis/Figures/vary_Pe_10_sigma.jpg}
    {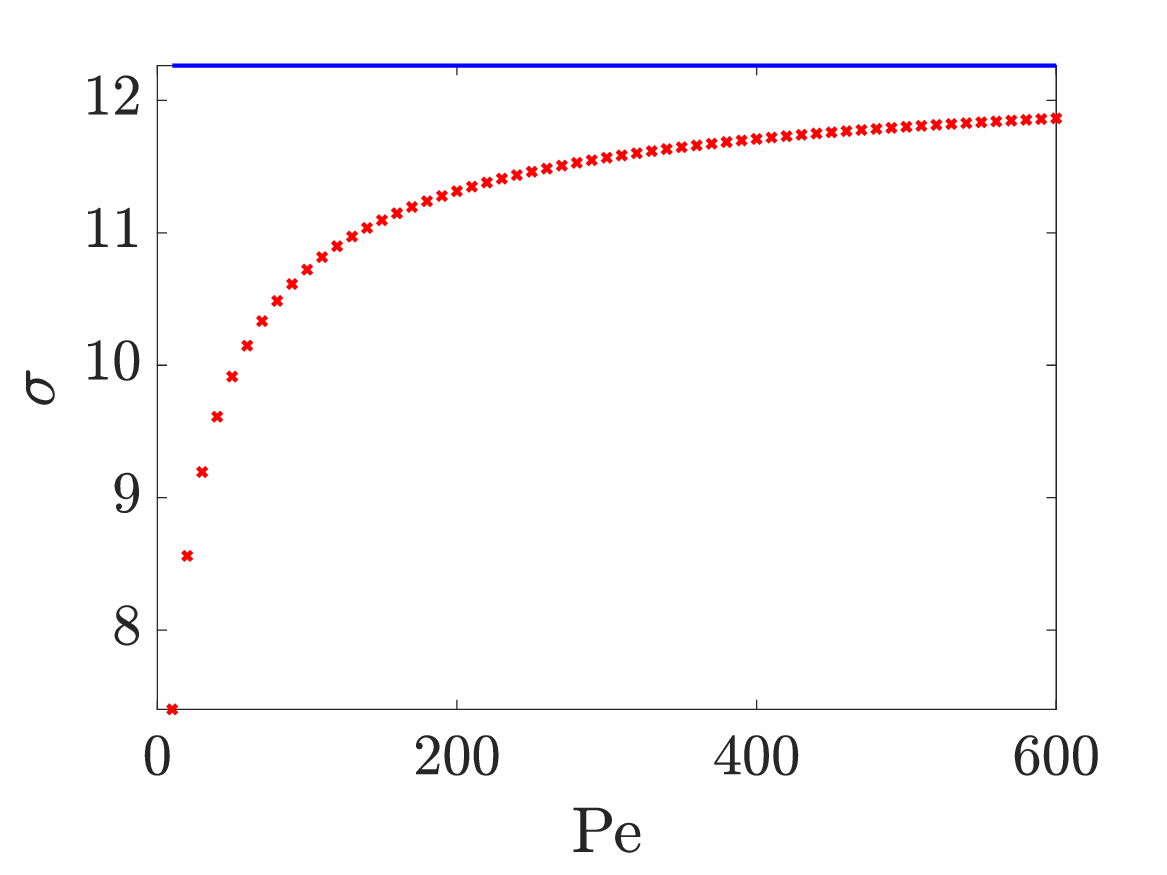}
    \caption{$\sigma$}
    \label{Fig:vary_Pe_sigma}
        \end{subfigure}
        \caption[First-order and second-order solution, with parameters $D = 2$, $\Ca = 1$, $\HR = 1$, and $\HC = 1$, varying $\Pe$.]{(a) Temperature $\theta$ and (b) aspect ratio $\phi$ at $x = 1$, along with (c) pulling tension $\sigma$, all plotted against P\'eclet number $\Pe$. First-order solution $\Pe \rightarrow \infty$ (blue curve) and second-order solution (red crosses). Other parameters: $D = 2$, $\Ca = 1$, $\HR = 1$, $\HC = 1$, $C=10$, $\phi_\text{in}=0.2$, $\theta_a=0.2990$.}
        \label{Fig:vary_Pe}
\end{figure}

Figure \ref{Fig:vary_Pe} shows the temperature $\theta$ and the aspect ratio $\phi$, both at $x=1$, as well as the pulling tension $\sigma$ for P\'eclet number in the range $10 \leq \Pe \leq 600$. Figure \ref{Fig:vary_Pe_theta} shows the temperature at $x = 1$, $\theta(1)$. The temperature solution of the second-order problem, $\theta_2(1)$ (red crosses), decreases as the P\'eclet number increases, approaching the solution of first-order problem ($\Pe \rightarrow \infty$), $\theta_1(1) = 0.3181$, as $\Pe$ becomes large. This shows that one actually needs a rather large value of the P\'eclet number for the conduction terms to become negligible.

Both the aspect ratio $\phi$ at $x = 1$ and pulling tension $\sigma$ of the second-order problem increase as P\'eclet number increases, approaching the solutions of the first-order problem ($\Pe \rightarrow \infty$), as shown in Figures \ref{Fig:vary_Pe_phi} and \ref{Fig:vary_Pe_sigma} respectively. The pulling tension needed in the case of the second-order problem is lower than for the first-order problem, because of the higher temperature throughout the glass arising from inclusion of conductive heat transport. With higher temperature, the glass is less viscous and hence requires less tension to achieve the draw ratio of $D = 2$. Figure \ref{Fig:vary_Pe} also show that for low P\'eclet number, there is a significant difference between the first-order and second-order solutions, indicating the importance of the inclusion of the second-order heat conduction term. 

%In conclusion, increasing the P\'eclet number for the second-order solution to $\Pe = 600$ results in the second-order solution approaching the first-order solution. The first-order model is obtained when we assume that the P\'eclet number is large enough. The second-order model approaching the first-order model as P\'eclet number increases gives us confidence that the second-order model is reasonable. The second-order temperature model is novel, and with this model we are able to find a more accurate solution than given by the first-order model for smaller P\'eclet number. \\

Note that Figures \ref{fig:Pe} and \ref{Fig:vary_Pe} have $\Pe = 600$ as the highest P\'eclet number, since this paper focuses on low draw ratio and  P\'eclet number applicable to pulled extrusion. However, solutions are readily obtained for larger P\'eclet number.  %This is shown in Table \ref{tab:varying_Pe}, where a P\'eclet number of 2500 approximates the first-order problem very well, with the temperature and aspect ratio match to 3 decimal places. However, we see that a P\'eclet number of 600 produces a solution that agrees with the first-order problem to 3 decimal places for the temperature and aspect ratio, hence we chose $\Pe = 600$ for Figures \ref{fig:Pe} and \ref{Fig:vary_Pe}. 

\section{Conclusion}
This paper focuses on the coupling of temperature with fluid flow in the modelling of preform extrusion, an important first stage in the process for fabrication of many micro-structured optical fibres. Temperature is known to be an important factor that affects geometry in fibre fabrication. However, it has not before been included in preform extrusion modelling. Further, while temperature has been included in previous modelling of the drawing of a preform to an optical-fibre, this work has assumed that heat conduction is negligible compared to advective heat transport \citep{JFM2019} because of the fast speed of the drawing process. 
%For the much slower extrusion process, we have clearly shown that axial heat conduction is important. 
Here we have examined the importance of axial heat conduction in the much slower extrusion process. % and shown it to be important. 
For simplicity, we have considered quasi-steady pulled extrusion of an axisymmetric tube. % by exploring negligible and non-negligible conduction.
Non-negligible conduction changes the temperature model from a first-order to a second-order differential equation. 
% The solution of the resulting system of ODEs requires an additional boundary condition. 
% %A steady leading-order model is derived using asymptotic expansions, utilising the slenderness of the preform. The steady state model in this paper is similar to the fibre drawing model of \cite{JFM2019}, which is an initial value problem that searches tension $\sigma$ to achieve a draw ratio $D = 2$. However, this paper considers both negligible and non-negligible heat conduction, changing the model to a second-order differential equation and requiring an additional condition. 
% The extra boundary condition chosen is a zero temperature gradient at the end of the domain, where the temperature approaches the ambient air temperature and the preform is sufficiently cold that it is effectively solid and deformation cannot occur. It is assumed that any error arising from the choice of this boundary condition is local and has no significant effect on the preform geometry. 

The extrusion problem with inclusion of heat conduction proves to be extremely numerically sensitive when standard numerical techniques are applied. 
%Unlike the straightforward problem in \cite{JFM2019} where the model is solved using a \textsc{Matlab} ordinary differential equation solver,
We hence developed a bespoke numerical procedure based on an iterative finite difference method. While forward-Euler finite-differencing was used to discretise the equations for most of the unknown variables, backward-Euler finite differencing was used for the differential equation for the temperature gradient. This was done to suppress non-physical exponential growth in the temperature and, we believe, is a novel solution method.  This strategy might be used in other mathematically similar problems. %The backward finite difference method also could be used in mathematically similar problems, a growing error in the boundary layer, to manage the error and give a physical solution. 

We have compared solutions of the system with a second-order temperature equation for different values of the P\'eclet number, $\Pe$. For $\Pe\rightarrow\infty$, our temperature equation reduces to first order and, in line with this, our solution to the second-order problem approaches that of the first-order problem as $\Pe$ became large, giving verification of our novel solution methodology. %approaches the first-order solution, giving us confidence that the second-order model is reasonable. 
%The second-order temperature model is novel, 
For small and even moderate P\'eclet number and a low draw ratio, we have demonstrated the importance of including heat conduction in the model. 

%{\color{red} Regarding this para on future work, discuss adapting the model and solution method to preforms of arbitrary geometry through use of the reduced time variable $\tau$ which is used in \citet{JFM2019}, rather than inclusion of heat conduction in the fibre draw model which probably has no real benefit?} 
Future work might consider adapting the model and solution method to extrusion of preforms of arbitrary geometry through use of the reduced time variable $\tau$, as used in \citet{JFM2019}. 
%Future work might consider the inclusion of heat conduction in the model of \citet{JFM2019} for drawing of a preform to a fibre, which involves both heating and cooling of the preform. 
This problem is expected to be amenable to solution using a similar numerical approach to that employed in this paper. Also, the inclusion of a temperature equation with both advective and conductive heat transport in unsteady gravitational extrusion is a topic for future exploration.

%{\color{red} I think this para on experimental work needs to discuss that $H_R$ and $H_C$ are not material parameters that can be measure but are parameters that need to be determined by comparison of model solutions with experiments and that this is an avenue for future work and, indeed, is needed to enable predictive modelling.} 
%Also, the parameter values used in this paper are chosen for exploratory purposes only. 
Finally we point out that the physical heat transfer parameter values $\beta$ and $h_w$, that appear in the dimensionless heat transfer parameters $\HR$ and $\HC$, are not quantities that can be directly measured experimentally. They %especially the heat transfer coefficients $\HR$ and $\HC$, can be obtained from experiments and then applied to this model. 
%Note that $\HR$ and $\HC$ 
are not material parameters but rather model parameters that depend on the physical extrusion setup. Both parameters need to be determined by comparison of model solutions with suitable experiments. This is an avenue for future work and, indeed, is needed to enable predictive modelling.

\section*{Credit authorship contribution statement}
\textbf{Eunice B. Yuwono}: Methodology, Software, Validation, Formal analysis, Investigation, Data curation, Writing - original draft, Visualisation. \textbf{Yvonne M. Stokes}: Conceptualization, Methodology, Software, Writing - review \& editing, Visualisation, Supervision, Funding acquisition, Project administration. \textbf{Jonathan J. Wylie}: Conceptualization, Writing - review \& editing. \textbf{Hayden Tronnolone}: Conceptualization, Writing - review \& editing, Supervision. 

\section*{Acknowledgements}
We would like to thank Professor Neville Fowkes, University of Western Australia, for his useful comments on this problem. YMS acknowledges the support of Australian Research Council grants LP200100541 and FT160100108, both of which provided funding for EBY. JJW was supported by the Research Grants Council of Hong Kong Special Administrative Region, China (CityU 11300720).

\addcontentsline{toc}{chapter}{Bibliography}

% What style to use and what file to use for the bibliography
%\bibliographystyle{elsarticle/elsarticle-num} 
\bibliographystyle{elsarticle-num-names} 
\bibliography{bibliography}

\end{document}